%% file: main.tex
\documentclass[sigconf,screen]{acmart}

\usepackage[utf8]{inputenc}
\usepackage{url}
\usepackage{subfig}
\usepackage{graphicx,multirow}
\usepackage{rotating}
\usepackage{colortbl}
\usepackage{multirow}
\usepackage{diagbox}
\usepackage{tablefootnote}
\usepackage{dcolumn}
\newcolumntype{d}[1]{D{.}{.}{#1}}

\usepackage[bottom]{footmisc}
\usepackage[ruled,vlined,linesnumbered]{algorithm2e}

\graphicspath{{./figures/},{./figures/plots/}}

\setcopyright{acmcopyright}
\copyrightyear{2023}
\acmYear{2023}

\acmSubmissionID{}

\begin{document}

\title{UEyes: An Eye-Tracking Dataset across User Interface Types}

\author{Yue Jiang}
\email{yue.jiang@aalto.fi}
\affiliation{%
  \institution{Aalto University}
  \country{Finland}
}

\author{Luis A. Leiva}
\email{name.surname@uni.lu} %
\affiliation{%
  \institution{University of Luxembourg}
  \country{Luxembourg}
}

\author{Paul R. B. Houssel}
\email{name.surname@uni.lu} %
\affiliation{%
  \institution{University of Luxembourg}
  \country{Luxembourg}
}

\author{Hamed R. Tavakoli}
\email{hamed.rezazadegan_tavakoli@nokia.com}
\affiliation{%
  \institution{Nokia Technologies}
  \country{Finland}
}

\author{Julia Kylmälä}
\email{julia.kylmala@aalto.fi}
\affiliation{%
  \institution{Aalto University}
  \country{Finland}
}

\author{Antti Oulasvirta}
\email{antti.oulasvirta@aalto.fi}
\affiliation{%
  \institution{Aalto University}
  \country{Finland}
}

\renewcommand{\shortauthors}{Y. Jiang et al.}

\begin{abstract} %
Different types of user interfaces differ significantly in the number of elements and how they are displayed. 
To examine how such differences affect the way users look at UIs, 
we collected and analyzed a large eye-tracking-based dataset, 
\textit{UEyes} (62 participants, 1,980 UI screenshots, near 20K eye movement sequences), 
covering four major UI types: webpage, desktop UI, mobile UI, and poster. 
Furthermore, we analyze and discuss the differences in important factors, such as color, location, and gaze direction 
across UI types, individual viewing strategies and potential future directions. This position paper is a derivative of our recent paper \cite{jiang2023ueyes} with a particular focus on the UEyes dataset.
\end{abstract}

\maketitle

\input{01-Introduction}

\input{02-Related-Work}

\input{03-Dataset}

\input{06-Discussion-and-Limitation}

\input{07-Conclusion}

\begin{acks}
We appreciate the active discussion with Yao (Marc) Wang.
This work was supported by Aalto University's Department of Information and Communications Engineering,
the Finnish Center for Artificial Intelligence (FCAI),
the Academy of Finland through the projects Human Automata (grant 328813) and BAD (grant 318559),
the Horizon 2020 FET program of the European Union (grant CHIST-ERA-20-BCI-001),
and the European Innovation Council Pathfinder program (SYMBIOTIK project, grant 101071147).
\end{acks}

\bibliographystyle{ACM-Reference-Format}
\bibliography{main}

\end{document}

%% file: 01-Introduction.tex
\section{Introduction}

The study of what captures users' attention when interacting with user interfaces (UIs) has been a longstanding area of interest in HCI research.
This understanding is essential for designers seeking to direct users' attention, communicate important information, and prevent visual overload~\cite{Rosenholtz11, Still10}.
Despite years of research on this topic, our understanding of how different types of UIs vary regarding visual saliency remains limited.
For example, posters usually feature only a few images, whereas desktop and mobile UIs tend to include more components, organized as widgets.
Understanding how these differences affect eye-movement patterns is essential.

In this paper, we present UEyes, a new eye-tracking dataset captured using a high-fidelity in-lab eye tracker on a large scale. While previous studies relied on mouse movements or manual annotations as proxies for eye movements, UEyes provides fine-grained ground-truth data on visual saliency. 
Our dataset includes around 20K multi-duration saliency maps and scanpaths of 62 users who viewed a total of 1,980 different UIs, 
comprising 495 UIs each from desktop, mobile, webpage, and poster applications.
Furthermore, we analyze and compare saliency-related tendencies across different types of UIs, addressing both bottom-up factors related to the visual primitives of the stimulus (e.g., color bias) and top-down (learned) factors related to the distribution of features in the dataset (e.g., location bias and scanpath direction). We present several previously unreported findings that illuminate what distinguishes particular UI types.

%% file: 02-Related-Work.tex
\section{Existing Visual Saliency Datasets}

Most visual saliency datasets available today focus on specific types of designs and provide a relatively small number of saliency results. These datasets are often limited to a particular type of visual design, with data collected from a small group of participants in a restricted context. For example, datasets such as MASSVIS~\cite{borkin2015beyond}, iSUN~\cite{xu2015turkergaze}, SALICON~\cite{jiang2015salicon}, MIT1003~\cite{judd2009learning}, MIT300~\cite{Judd2012benchmark}, NUSEF~\cite{ramanathan2010eye}, and Leiva et al.~\cite{leiva2020understanding} typically encompass only one specific type of image, such as indoor and outdoor natural images, mobile user interfaces, visual flows in comics, webpages, and posters.
While CAT2000~\cite{borji2015cat2000} includes 20 categories, all of them are classes of natural images, with additional augmented natural images such as sketches and cartoons, and noisy natural images like low-resolution scenes and Gaussian-noised images. However, saliency prediction models trained on specific image types are often limited in their ability to adapt to broader types of images and cannot generalize to predict visual saliency for a wide range of them.
To address this limitation, we collected the UEyes dataset for this work, which includes eye-tracking data for four common categories of user interfaces and a wide variety of images, with a focus on diverse visual designs. By including a broad range of UIs, our dataset provides a valuable resource for researchers seeking to develop saliency prediction models that can adapt to different types of images and generalize across various domains.

While prior research has explored the use of crowdsourcing to collect saliency-related data (e.g., Imp1k~\cite{fosco2020predicting} and SALICON~\cite{jiang2015salicon}), these methods cannot provide the high-fidelity in-lab eye-tracking data that is crucial for accurate results. Proxy sensors, such as webcams and cursor movements, present their own issues, including accuracy issues with webcam-based methods~\cite{xu2015turkergaze} during facial landmark tracking, eye region extraction, and calibration, and slower, more deliberative cognitive processes with cursor-based approaches~\cite{bednarik2007validating, jiang2015salicon, kim2015crowdsourced, kim2017bubbleview} than those involved in eye movements.
To address these limitations, we collected real-time eye-tracking data via an eye tracker and focused on common UI types to systematically analyze and compare user viewing behavior.

%% file: 03-Dataset.tex
\section{Dataset Collection}

The UEyes dataset is composed of both the 1,980 UI screenshots and the associated metadata and eye-tracking logs 
from 62 viewers, collected in a laboratory by means of a modern eye tracker. 
This dataset contains 495 screenshots from each of the following UI types:
\begin{description}

\item[Webpage:] We collected 494 webpage images from the Alexa 500 dataset~\cite{Alexa500}, 1,507 images from the Visual Complexity and Aesthetics dataset~\cite{miniukovich2020relationship}, and 200 images from the Imp1k dataset~\cite{fosco2020predicting}. We extended the breadth of the webpage image set by capturing 103 additional webpage screenshots.
\item[Desktop UI:] The desktop UI image set contains the Waltteri Github desktop UI dataset~\cite{desktopUI}, representing 51 desktop UIs, and an additional 303 desktop UI images collected in line with the criteria presented below.
\item[Mobile UI:] We extracted a sample of 1,761 images from among the 46,064 mobile UI images from the RICO dataset~\cite{deka2017rico}. 
We extended the set with 42 further mobile UI images.
\item[Poster:] The poster image set contains 200 ads and 198 infographics from the Imp1k dataset~\cite{fosco2020predicting}, along with 103 additional posters we collected.
\end{description}

To ensure a diverse and representative dataset, we added new images that were either substantially different from those in the pre-existing dataset or commonly used in day-to-day life. For mobile UI images, we collected apps from various categories, including school, library, music, and settings. We also added more desktop UI images to create a balanced final dataset. We filtered out any images containing pornography and randomly sampled images of each type to create 55 "image blocks" for user assessment. Each block consisted of nine images representing each UI type, for a total of 36 images per block. To mimic users' typical viewing experience, we adjusted the screen angle for each participant during the data collection process. We used the same visual angle for all UI types, including mobile UIs, to ensure a fair comparison. This consistent presentation across types prevented the tracking technology's accuracy limits from disproportionately affecting the mobile UI results, enabling us to conduct consistent data collection and analysis across all UI types.

\paragraph{\textbf{Participants}}
We recruited a total of 66 participants, comprising 23 males and 43 females, through mailing lists and social media promotions. The average age was 27.25 (SD\ =\ 7.26). All participants had normal vision, with 43 of them having uncorrected vision and 23 wearing glasses or contact lenses for corrected-to-normal vision. None of the participants were colorblind.
We excluded the gaze data of four participants due to inaccurate eye-tracking calibration. Each user participated in the study for an hour and received 30 EUR as compensation for their time and effort.

\paragraph{\textbf{Experimental Design}}
Our system selected nine blocks randomly from a pool of 55 blocks for each user, resulting in a total of 324 images shown to each user. We tried to make sure each block has been shown to about the same number of users. Each block contained nine images for each UI type. The order of presentation for the images within each block was randomized.

\paragraph{\textbf{Apparatus}}
We presented the images on a 24-inch HP Compaq LA2405wg desktop monitor with dimensions of $32.5\times 52$ cm and a resolution of $1920 \times 1200$ px. To collect high-quality gaze data, we used a Gazepoint GP3 eye tracker with a sampling rate of 60 Hz. The eye tracker was positioned under the screen and tilted upwards, with its angle adjusted for each participant. The eye-tracking software, Gazepoint Control, provided feedback to ensure that the participant was seated at a distance of approximately 50--65 cm from the tracker.

\paragraph{\textbf{Procedure}}
The tracker was calibrated using Gazepoint Control's nine-point calibration and tested on the calibration test screen. To ensure the quality of the data in the post-processing stage, each participant was then shown three images of grids with different sizes and instructed to look at the corners of the grids, starting from the top left and moving clockwise. Following the calibration, each participant completed nine blocks with self-managed breaks. In each block, the participant was presented with UI images and asked to examine them for seven seconds, imagining they were in a corresponding real-world situation. No specific task was assigned, following the methodology of other bottom-up saliency studies. After the last block of UI images, the participant filled out a demographics questionnaire.

\setlength{\tabcolsep}{1.5pt}
\def\arraystretch{0.3}%
\begin{figure*}[h!]
\def\w{0.16\linewidth}
 \centering
\begin{tabular}{c *6c}
& \bf Input & \bf Saliency Maps & \multicolumn{4}{c}{\bf Scanpaths} \\
\bf \begin{turn}{90} 
\bf \ \ \ \ \ Webpage
\end{turn} &
\includegraphics[width=\w]{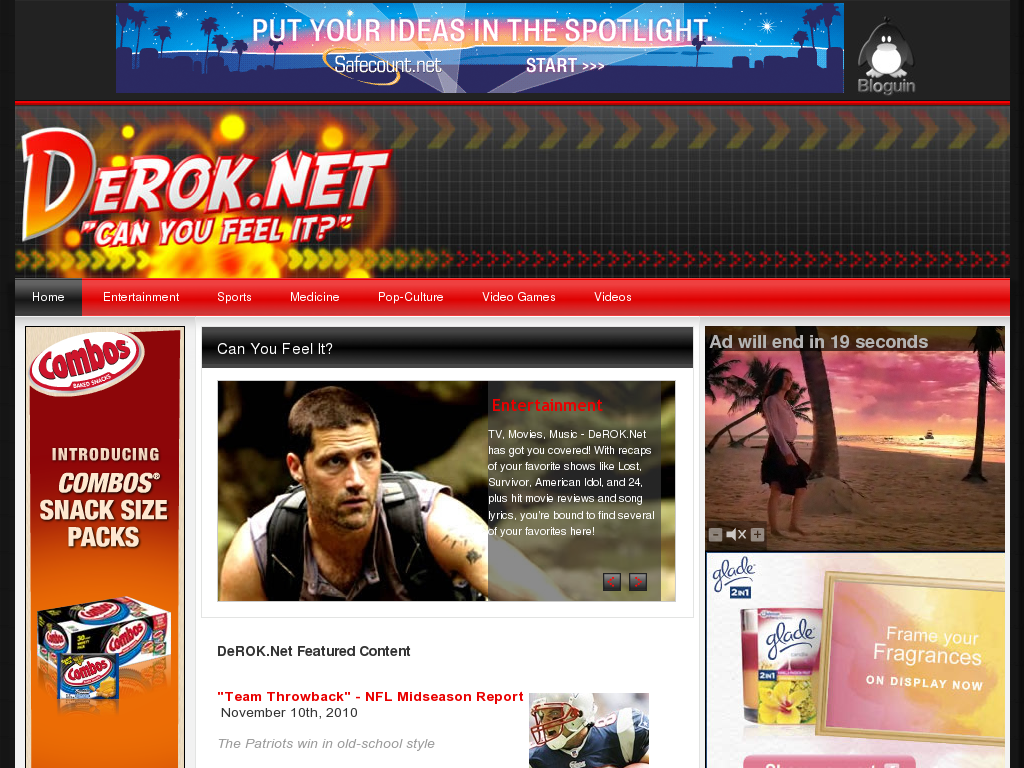} &
\includegraphics[width=\w]{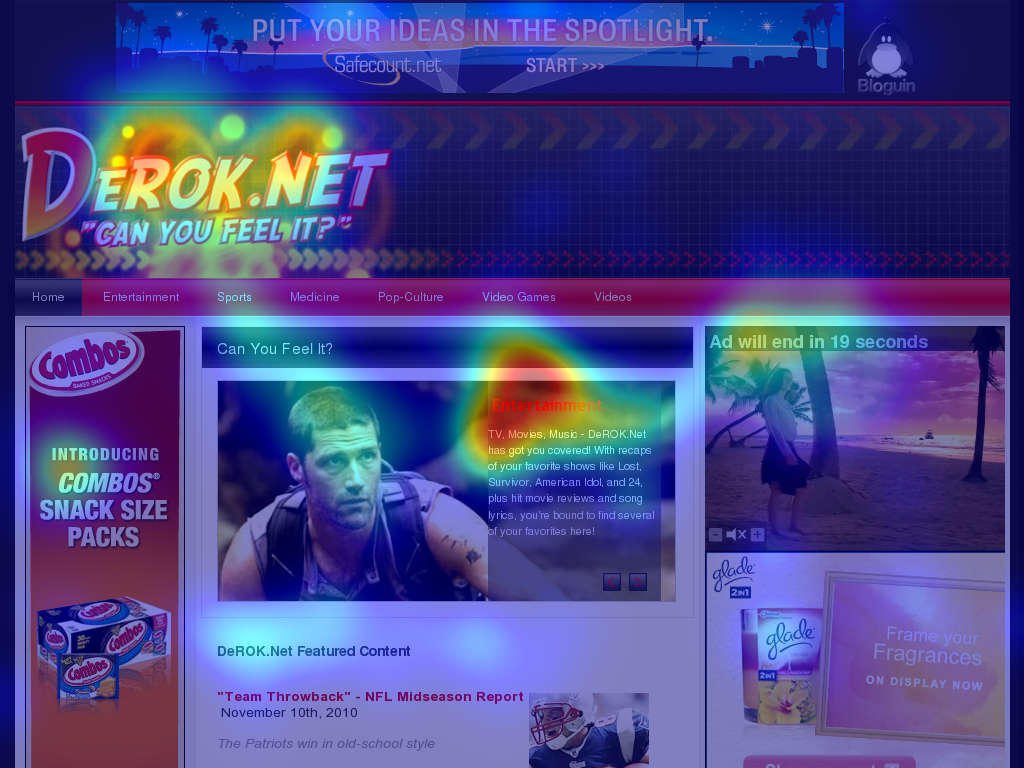} &
\includegraphics[width=\w]{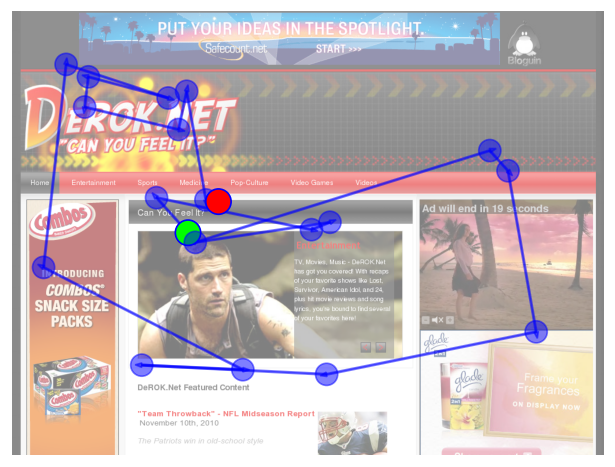} &
\includegraphics[width=\w]{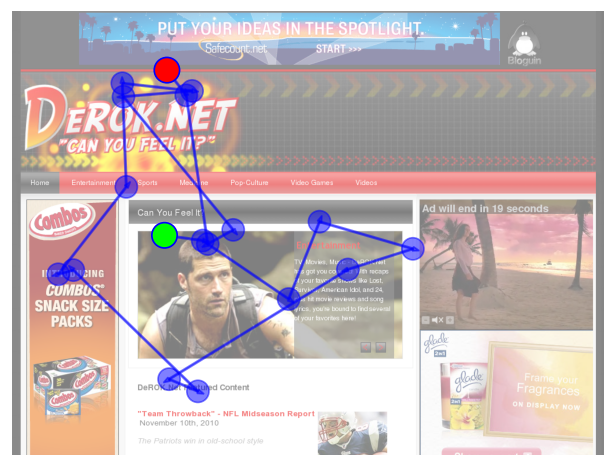} &
\includegraphics[width=\w]{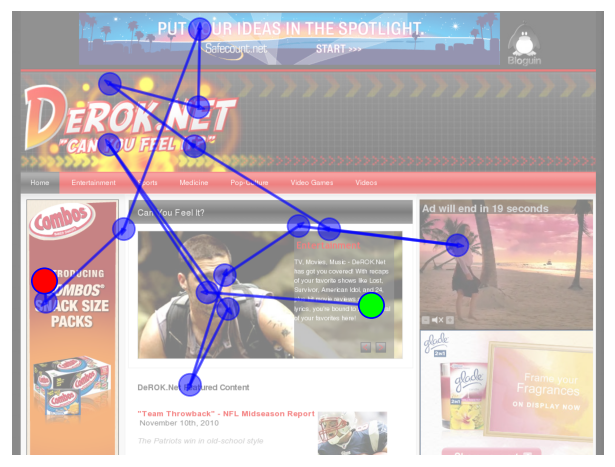} &
\includegraphics[width=\w]{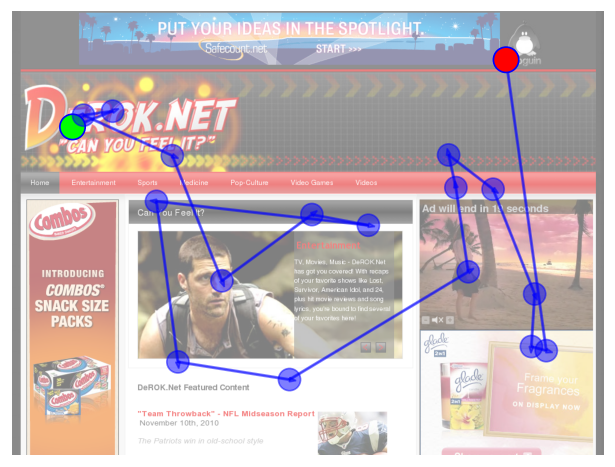}
\\
\bf \begin{turn}{90} 
\bf \ \ \ \   Webpage 
\end{turn} &
\includegraphics[width=\w]{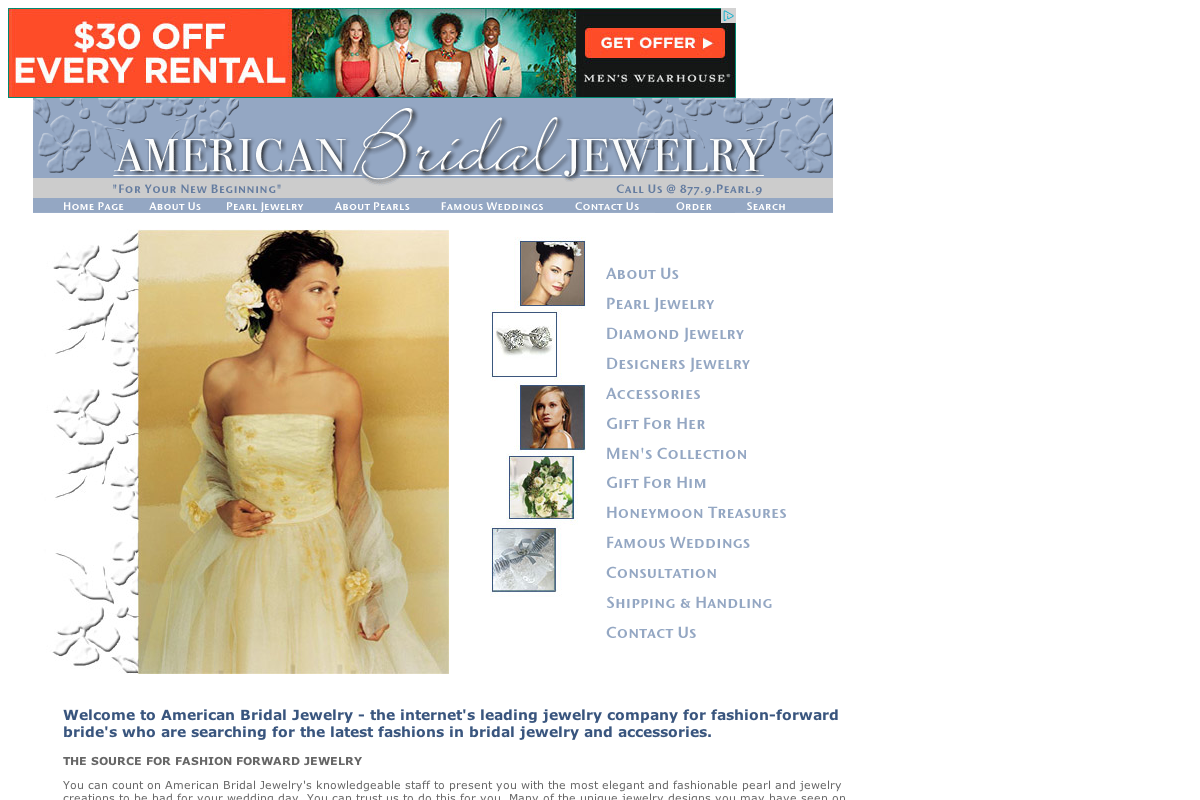} &
\includegraphics[width=\w]{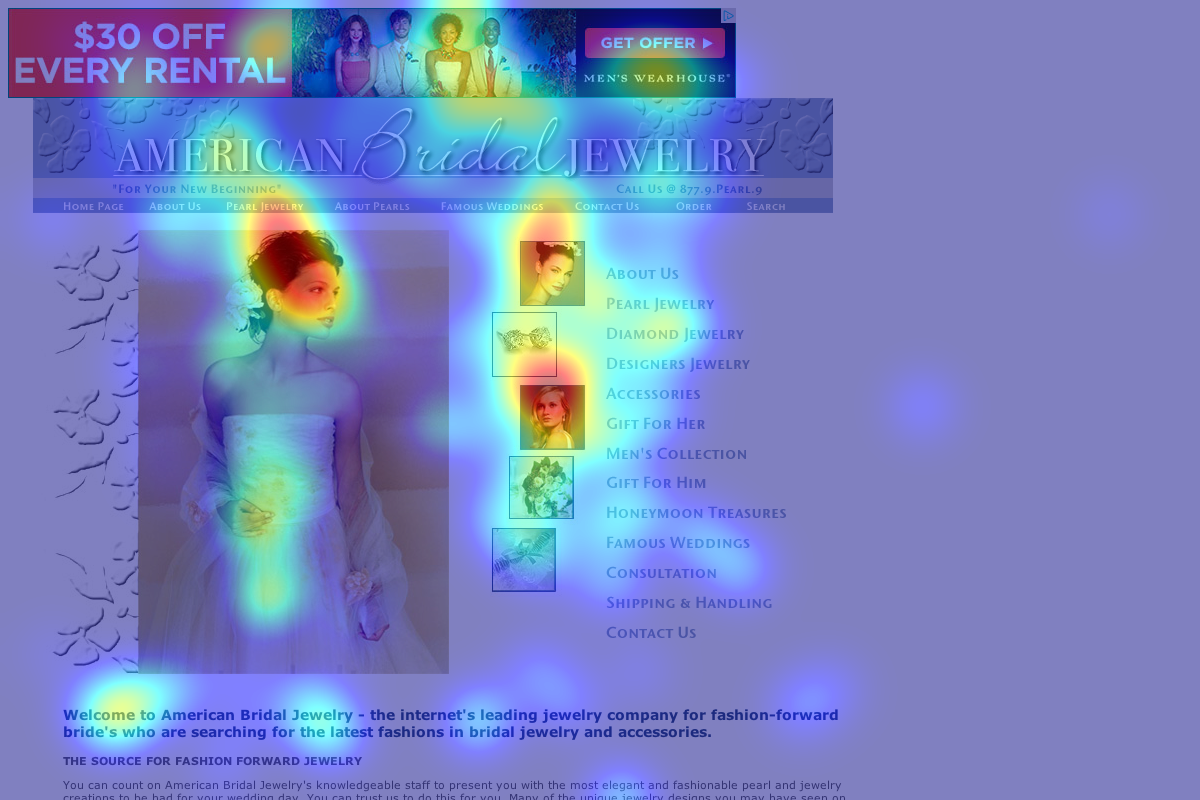} &
\includegraphics[width=\w]{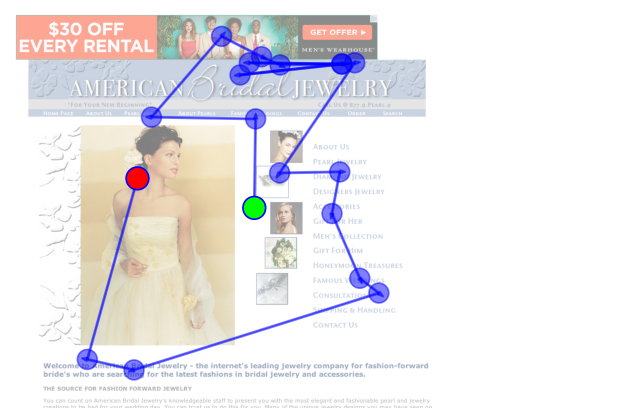} &
\includegraphics[width=\w]{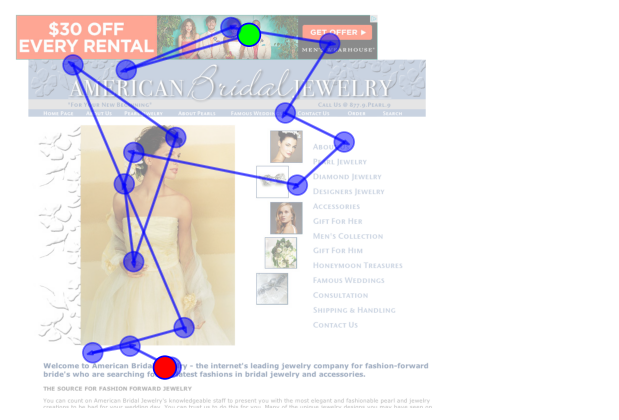} &
\includegraphics[width=\w]{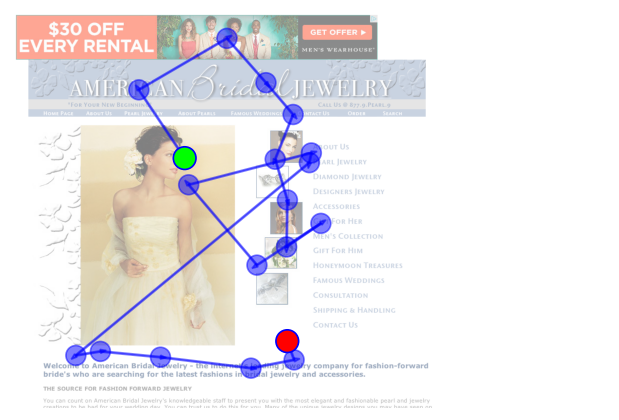} &
\includegraphics[width=\w]{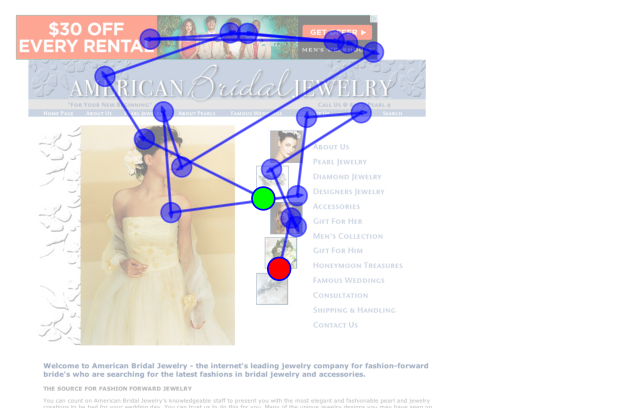}
\\
\bf \begin{turn}{90} 
\bf Desktop UI
\end{turn} &
\includegraphics[width=\w]{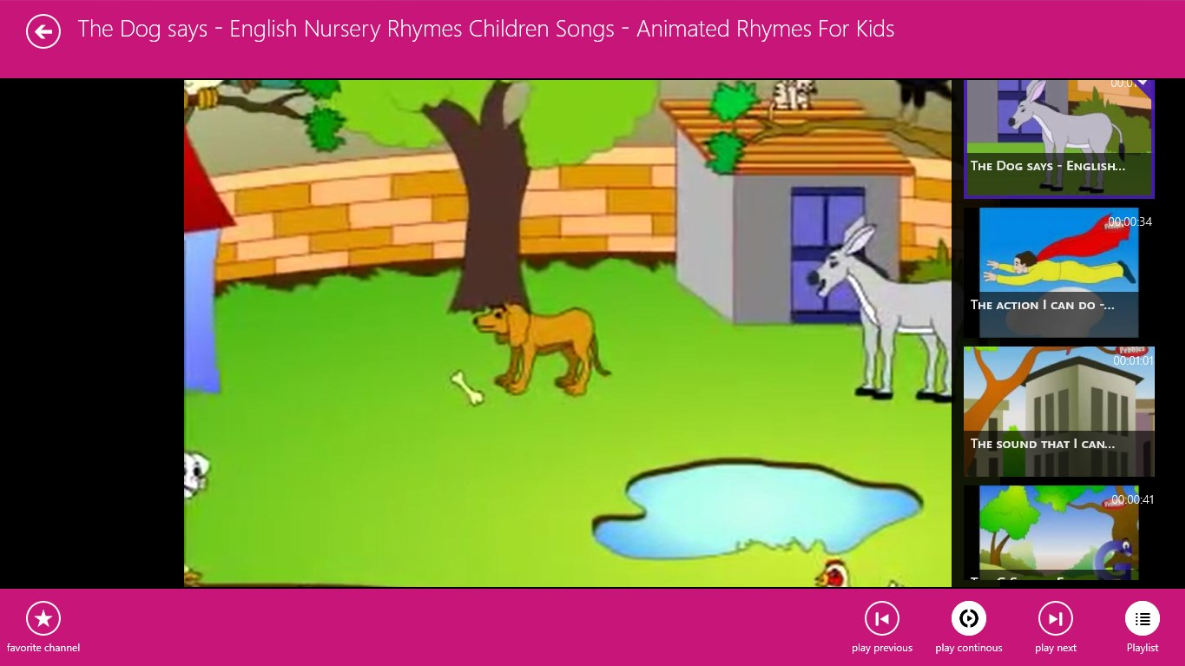} &
\includegraphics[width=\w]{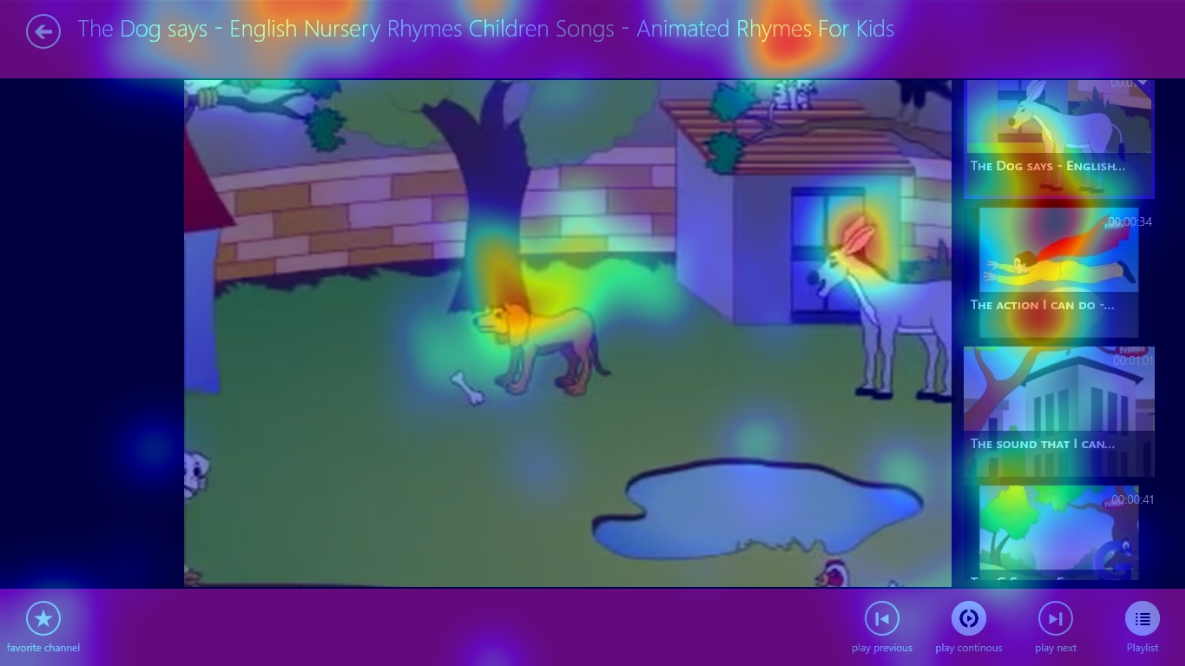} &
\includegraphics[width=\w]{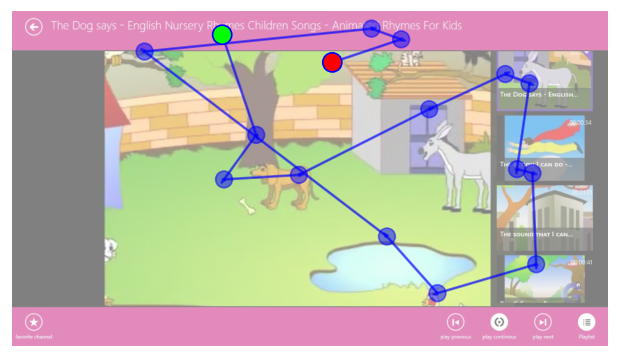} &
\includegraphics[width=\w]{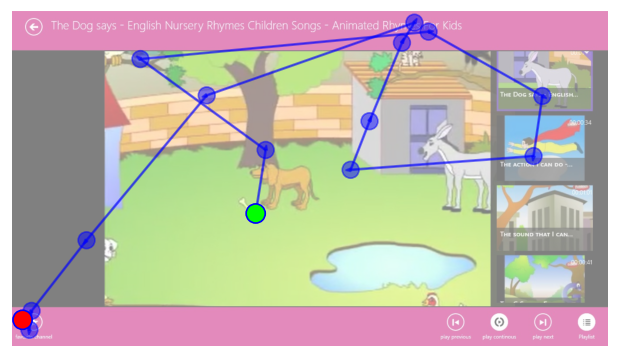} &
\includegraphics[width=\w]{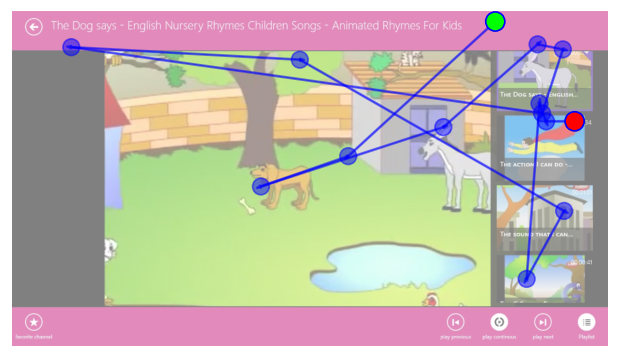} &
\includegraphics[width=\w]{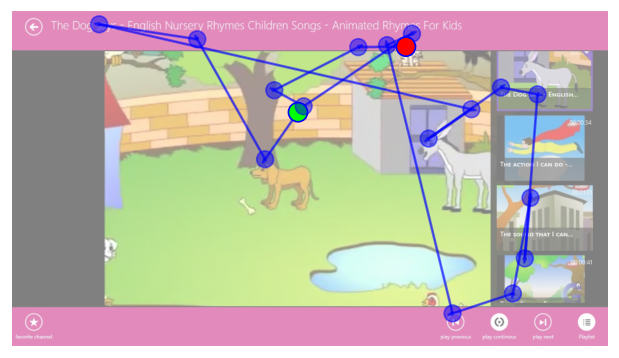}
\\
\bf \begin{turn}{90} 
\bf Desktop UI
\end{turn} &
\includegraphics[width=\w]{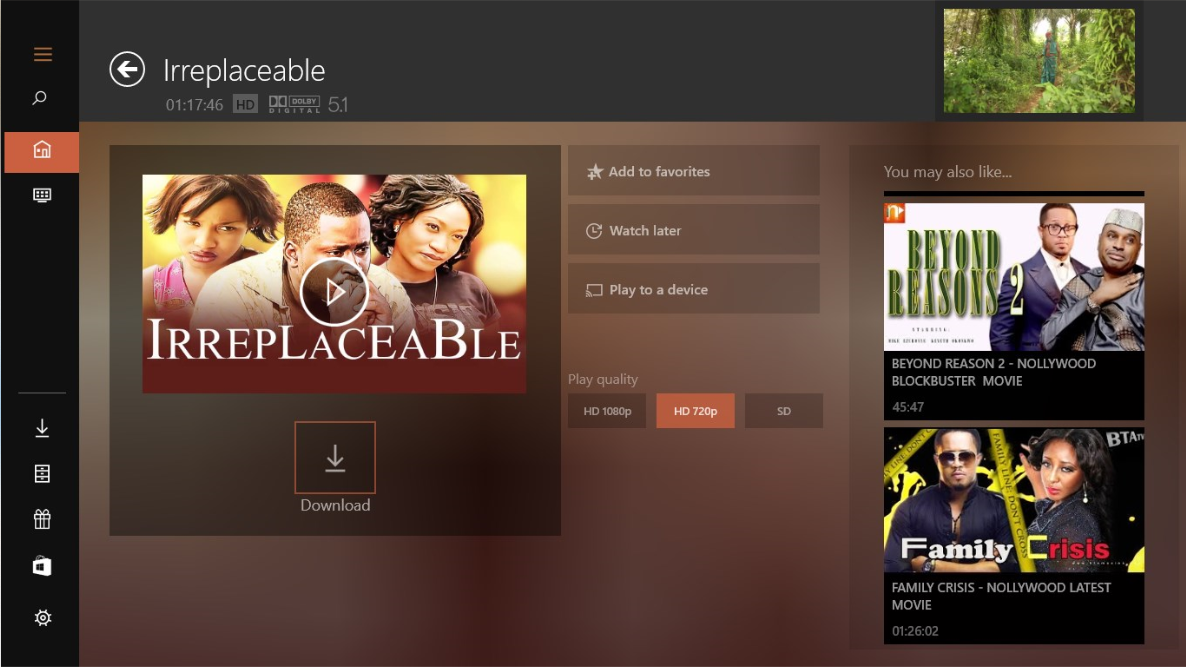} &
\includegraphics[width=\w]{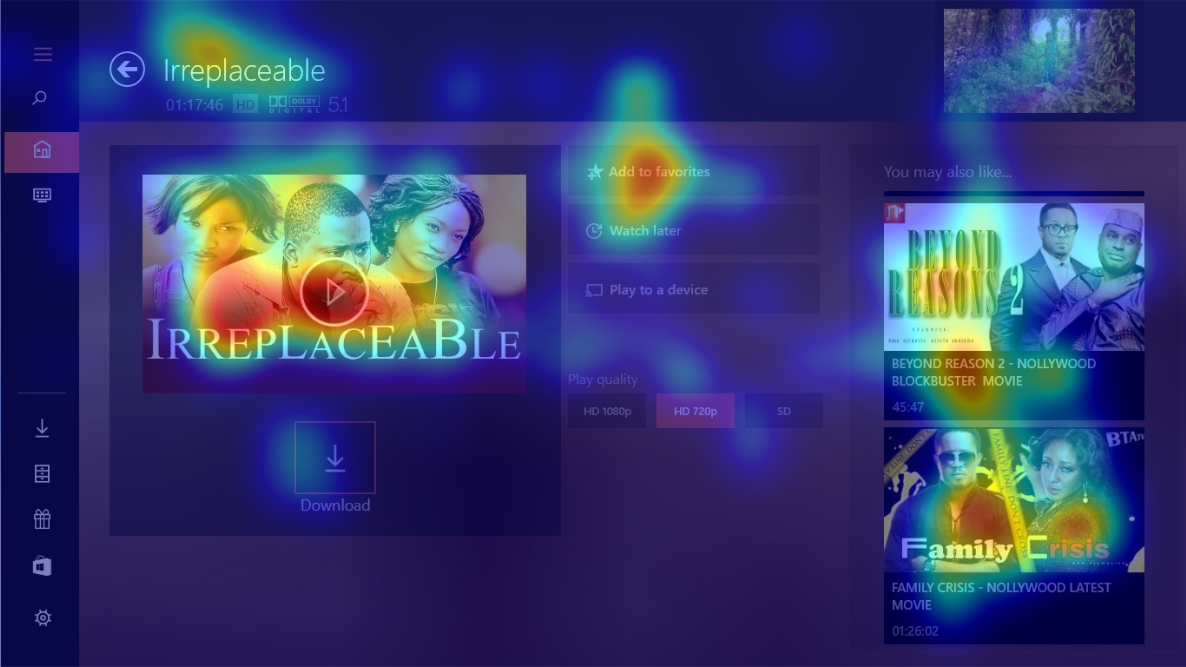} &
\includegraphics[width=\w]{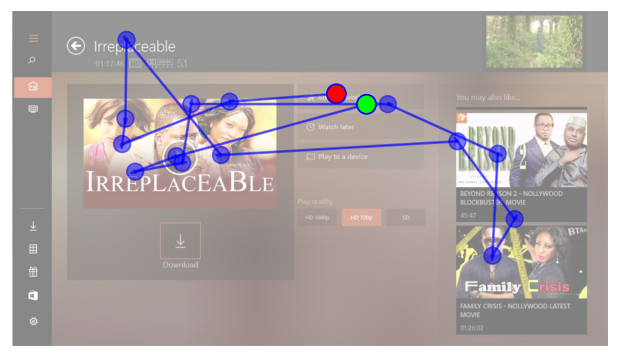} &
\includegraphics[width=\w]{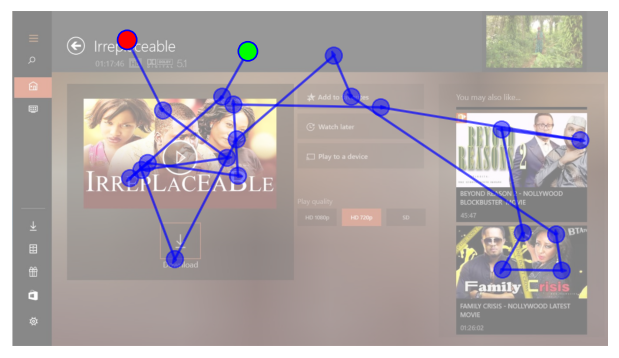} &
\includegraphics[width=\w]{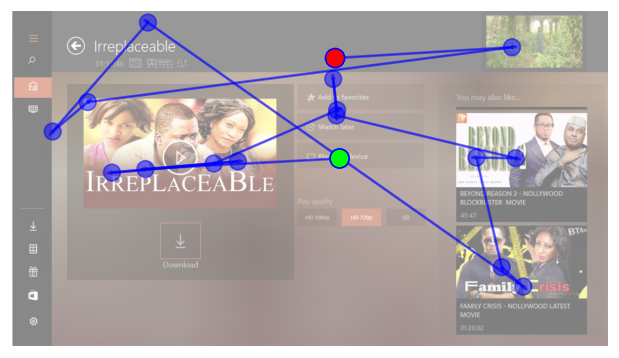} &
\includegraphics[width=\w]{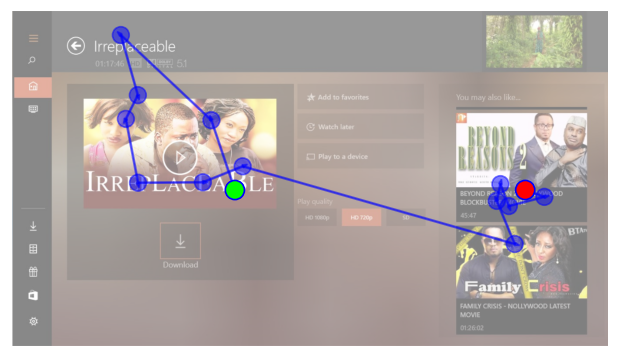}
\\
\bf \begin{turn}{90} 
\bf \ \ \ \ \ \ \ \ \ \ Mobile UI
\end{turn} &
\includegraphics[width=0.103\linewidth]{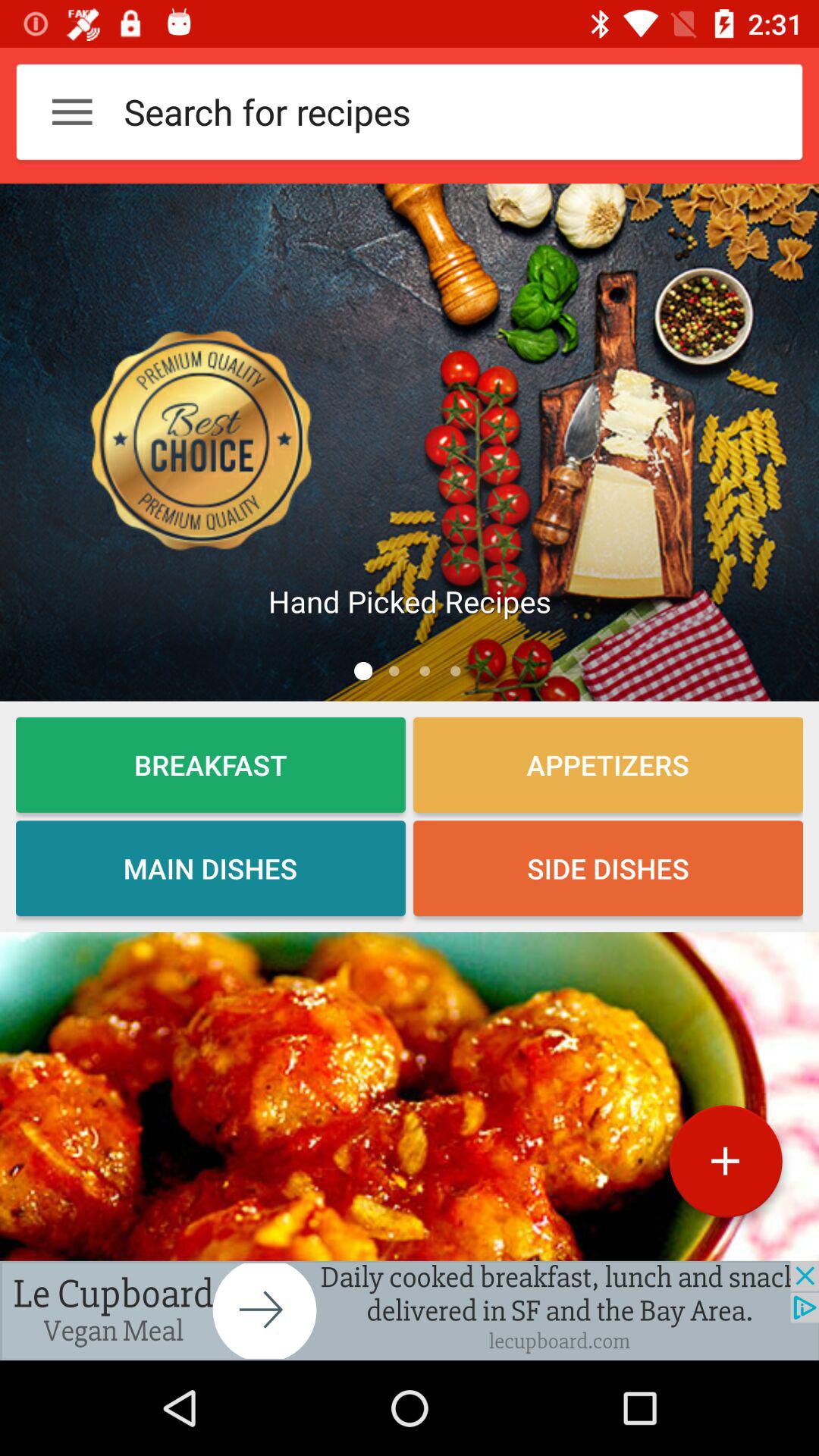} &
\includegraphics[width=0.103\linewidth]{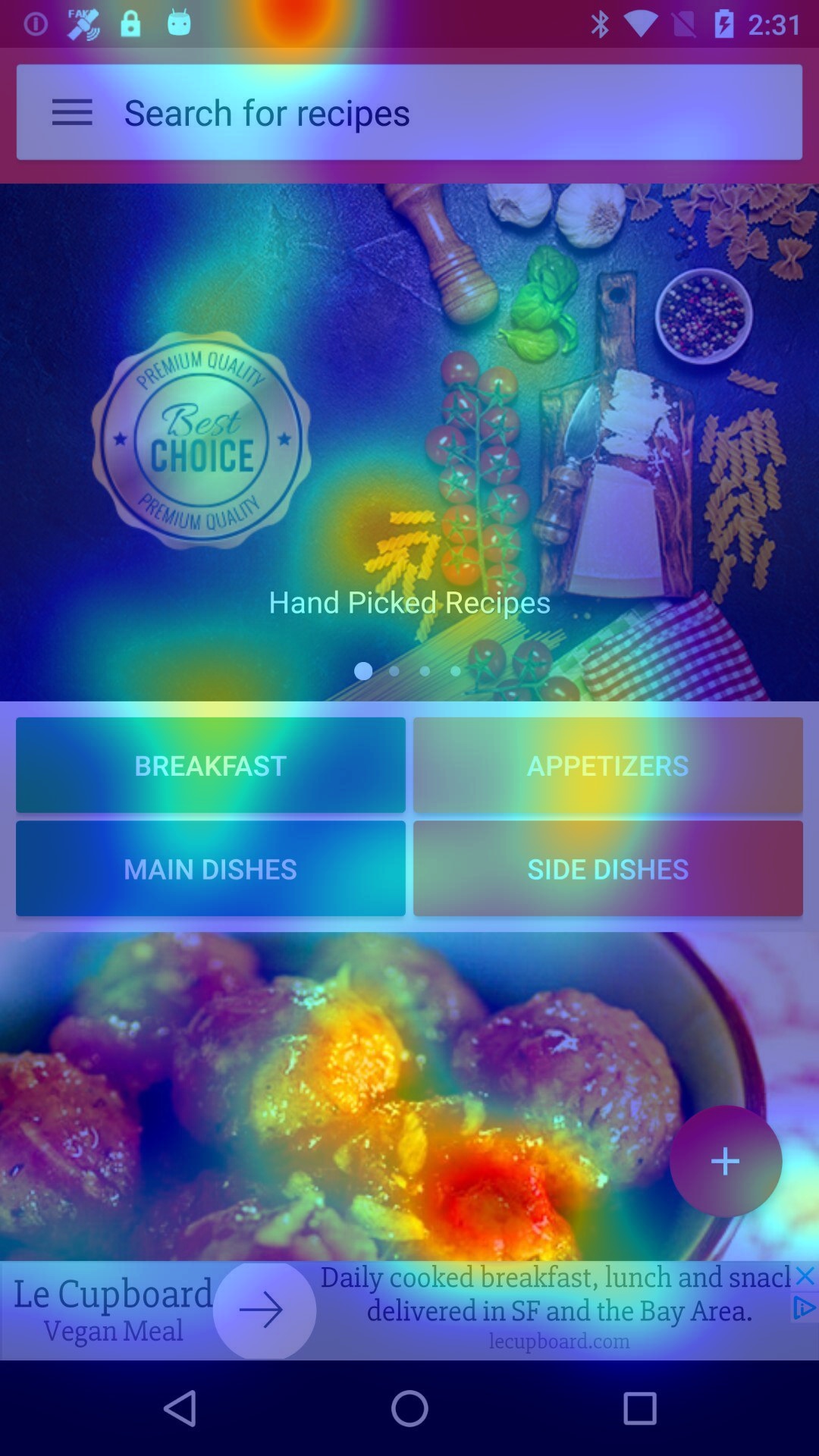} &
\includegraphics[width=0.11\linewidth]{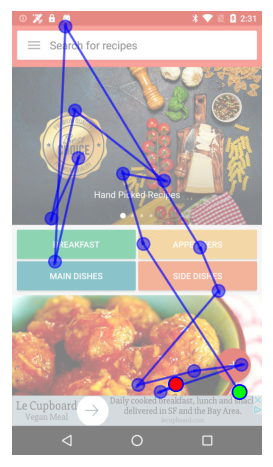} &
\includegraphics[width=0.11\linewidth]{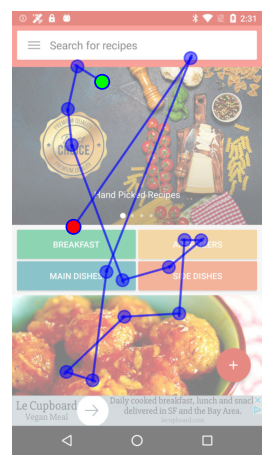} &
\includegraphics[width=0.11\linewidth]{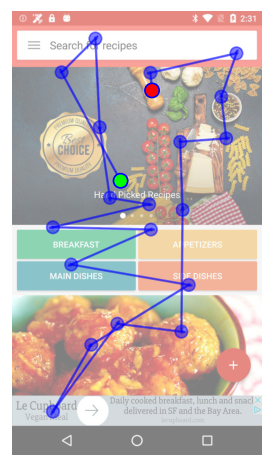} &
\includegraphics[width=0.11\linewidth]{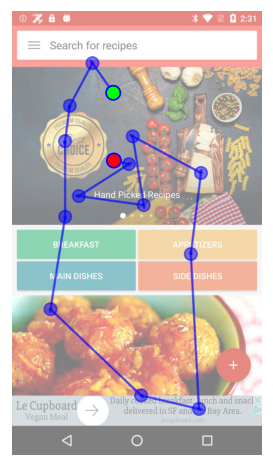}
\\
\bf \begin{turn}{90} 
\bf \ \ \ \ \ \ \ \ \ \ \ \ \ Mobile UI
\end{turn} &
\includegraphics[width=0.103\linewidth]{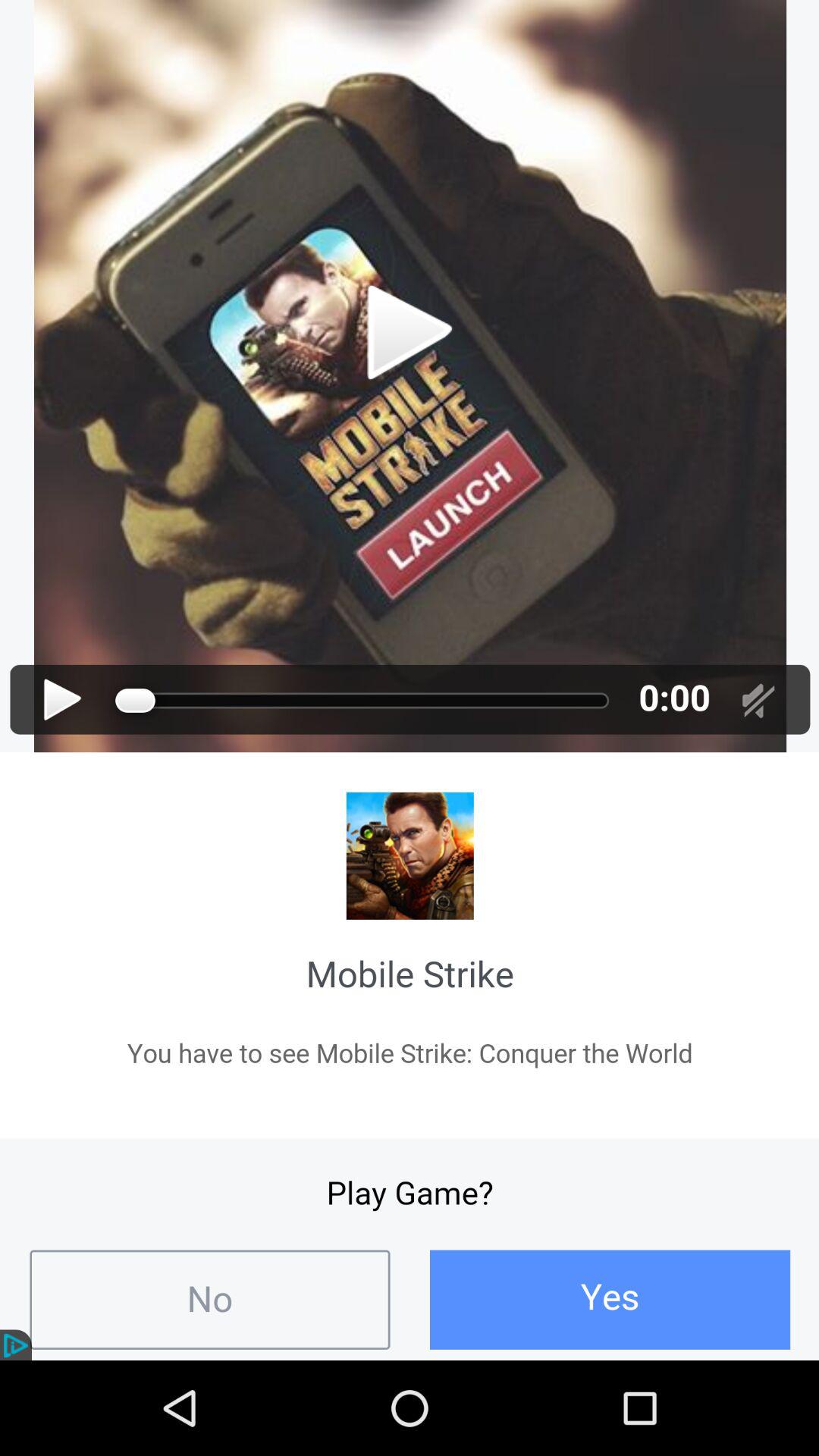} &
\includegraphics[width=0.103\linewidth]{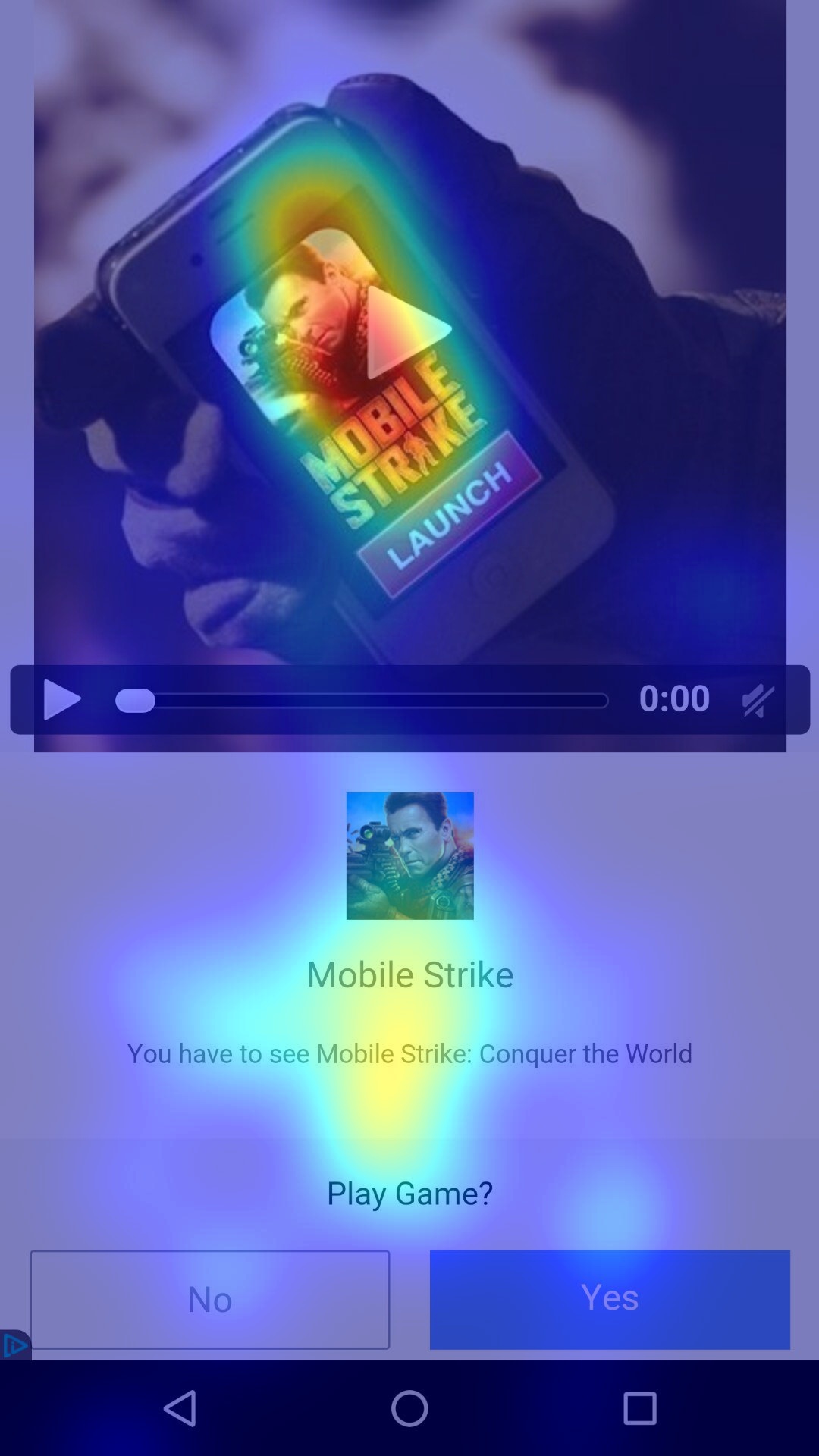} &
\includegraphics[width=0.11\linewidth]{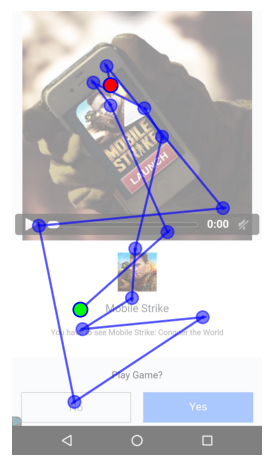} &
\includegraphics[width=0.11\linewidth]{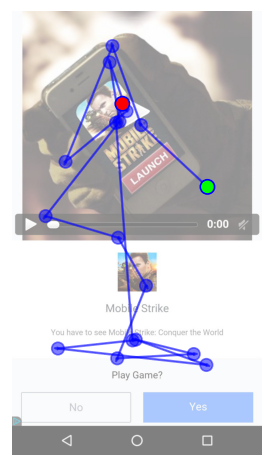} &
\includegraphics[width=0.11\linewidth]{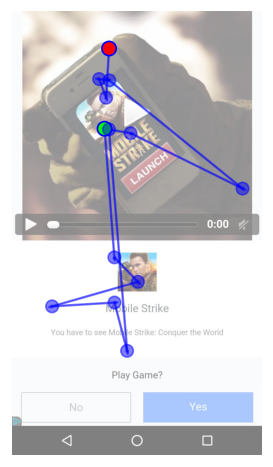} &
\includegraphics[width=0.11\linewidth]{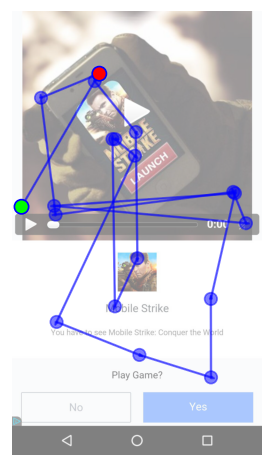}
\\
\bf \begin{turn}{90} 
\bf \ \ \ \ \ Poster
\end{turn} &
\includegraphics[width=\w]{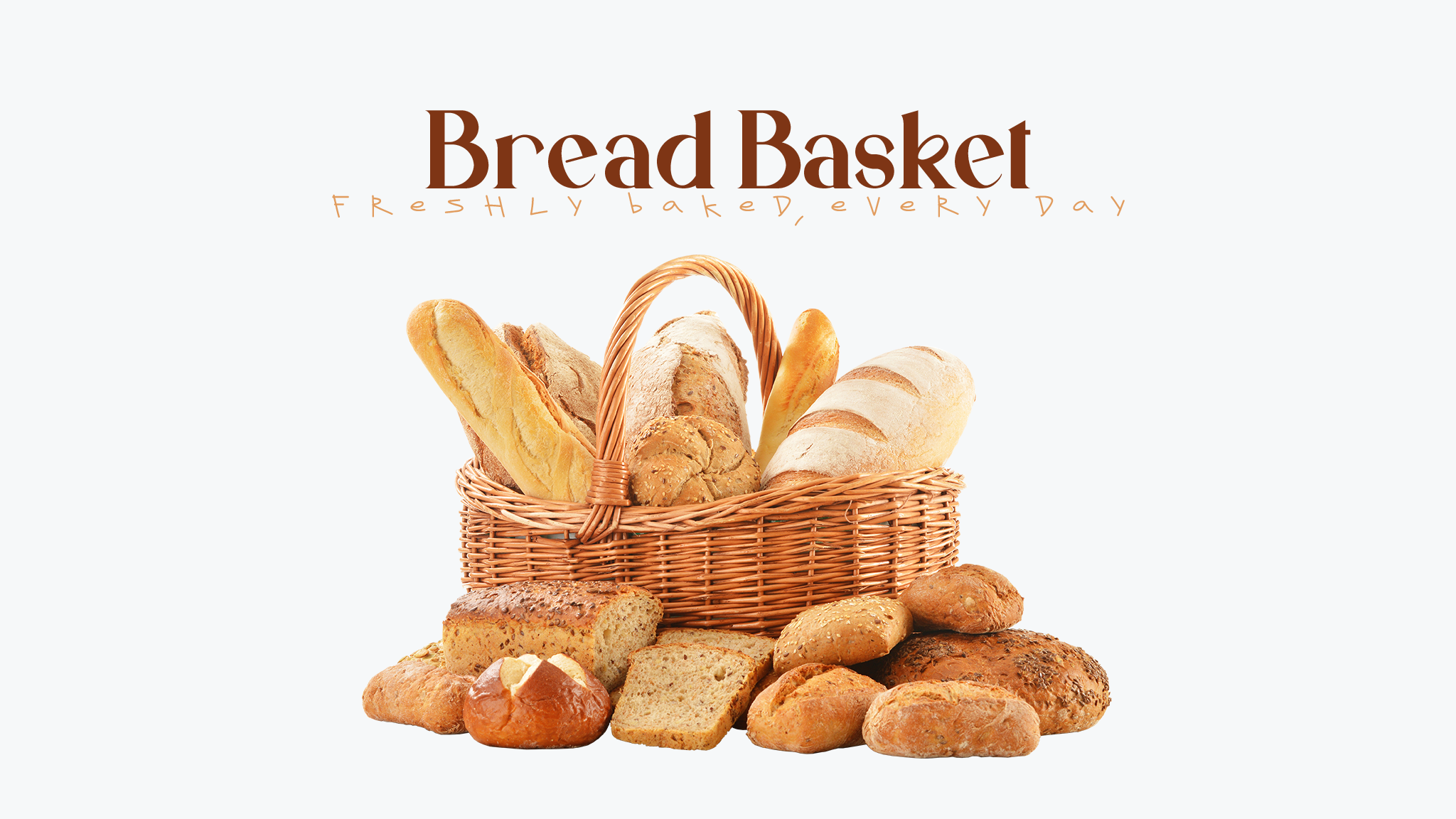} &
\includegraphics[width=\w]{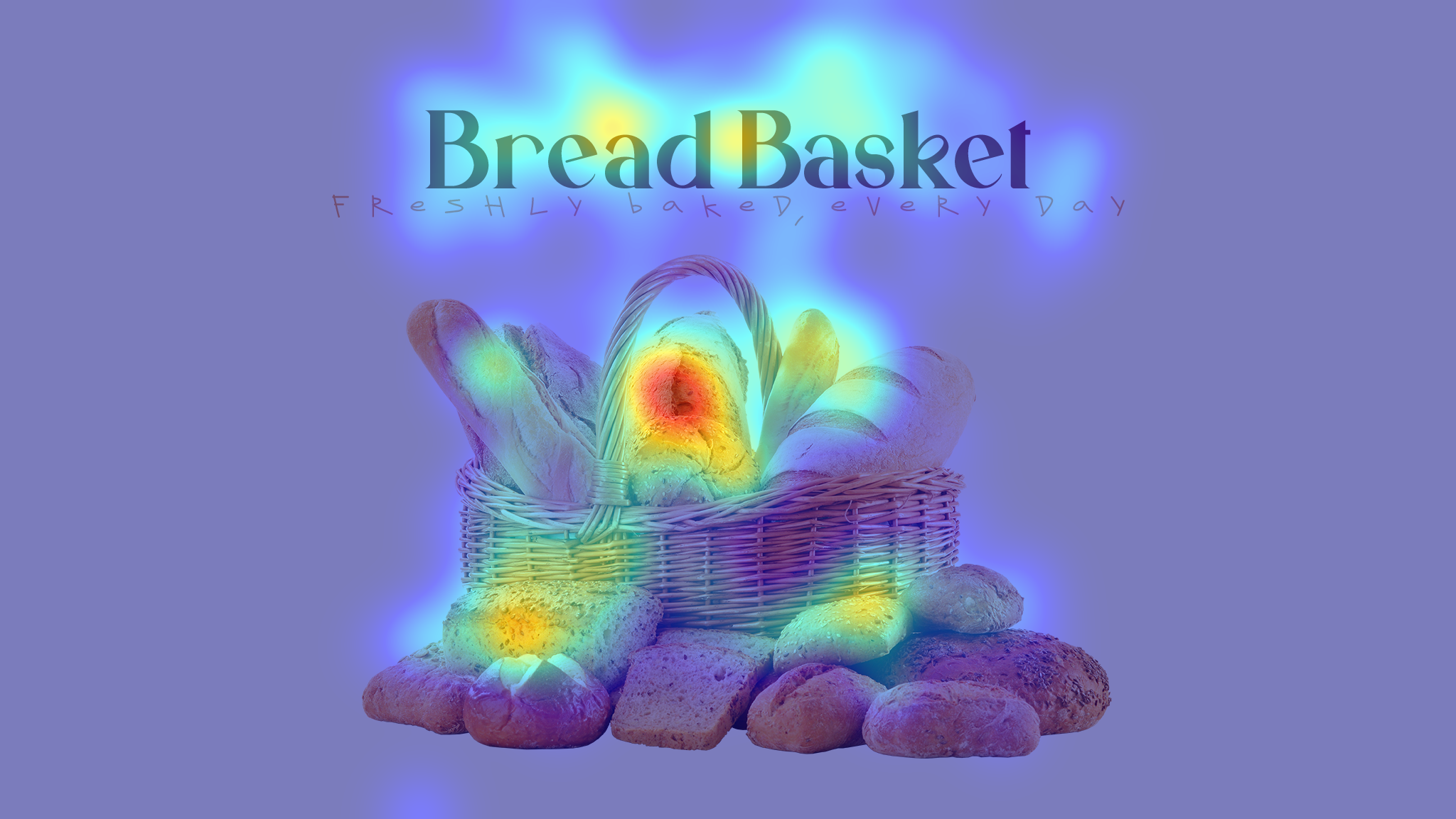} &
\includegraphics[width=\w]{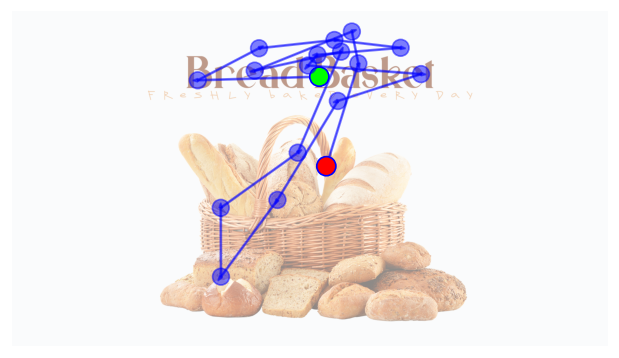} &
\includegraphics[width=\w]{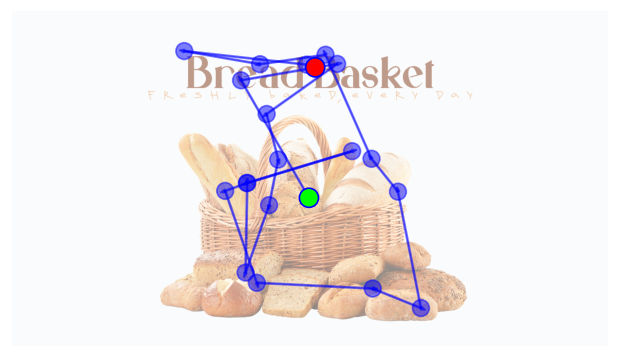} &
\includegraphics[width=\w]{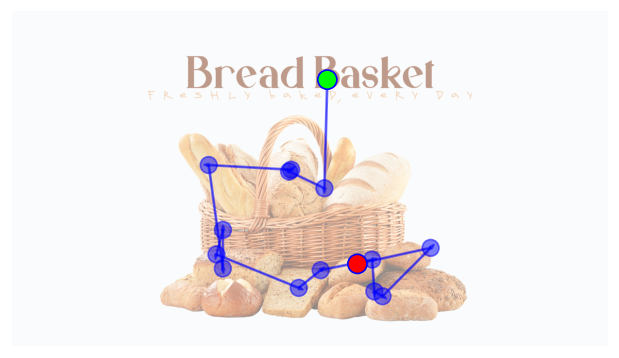} &
\includegraphics[width=\w]{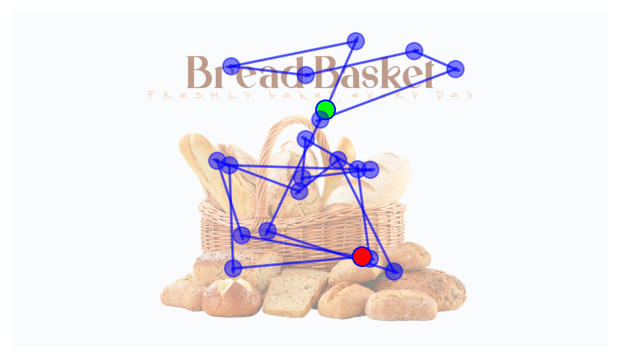}
\\
\bf \begin{turn}{90} 
\bf \ \ \ \ \ \ \ \ \ \ \ \ Poster
\end{turn} &
\includegraphics[width=0.103\linewidth]{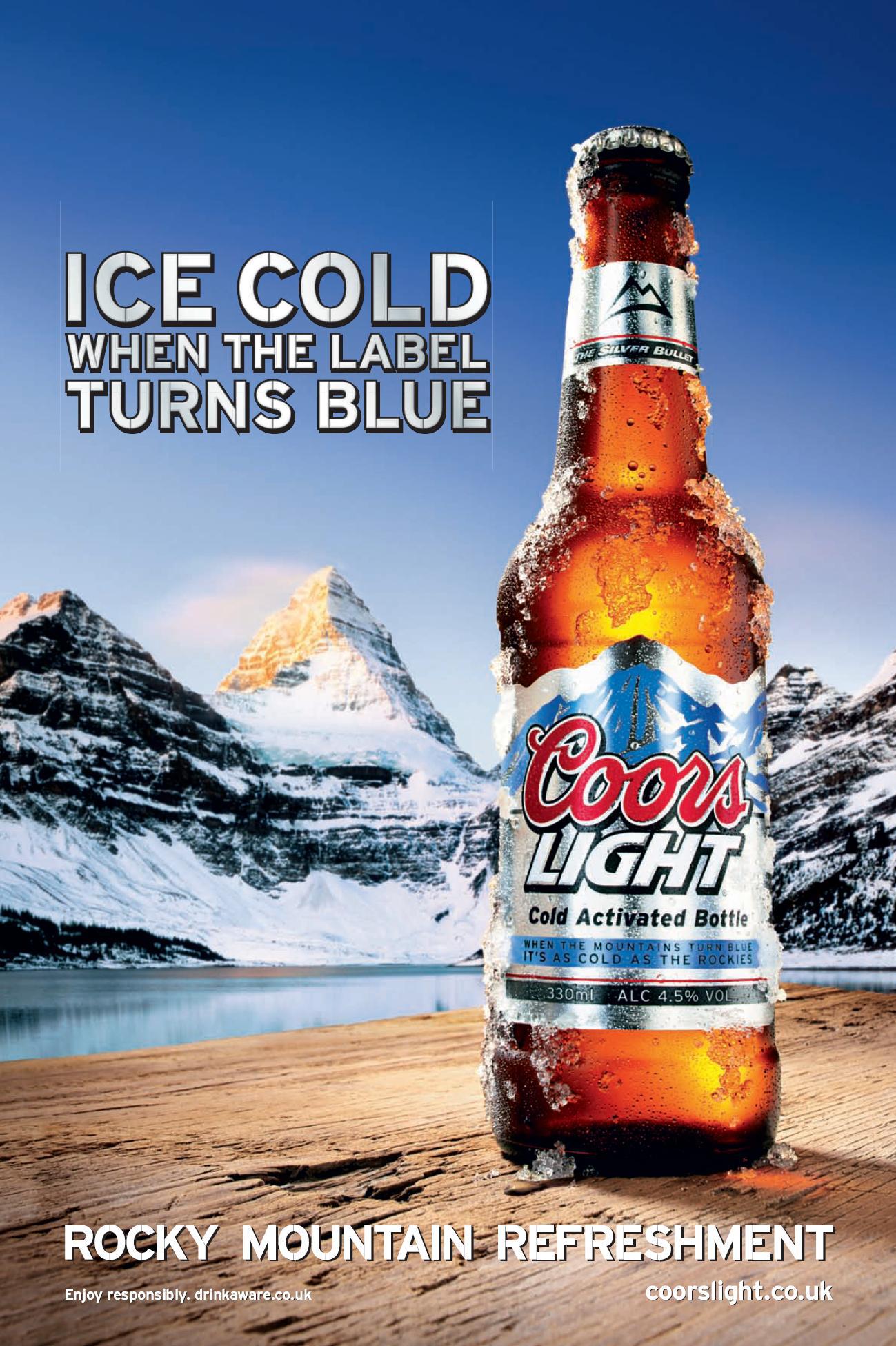} &
\includegraphics[width=0.103\linewidth]{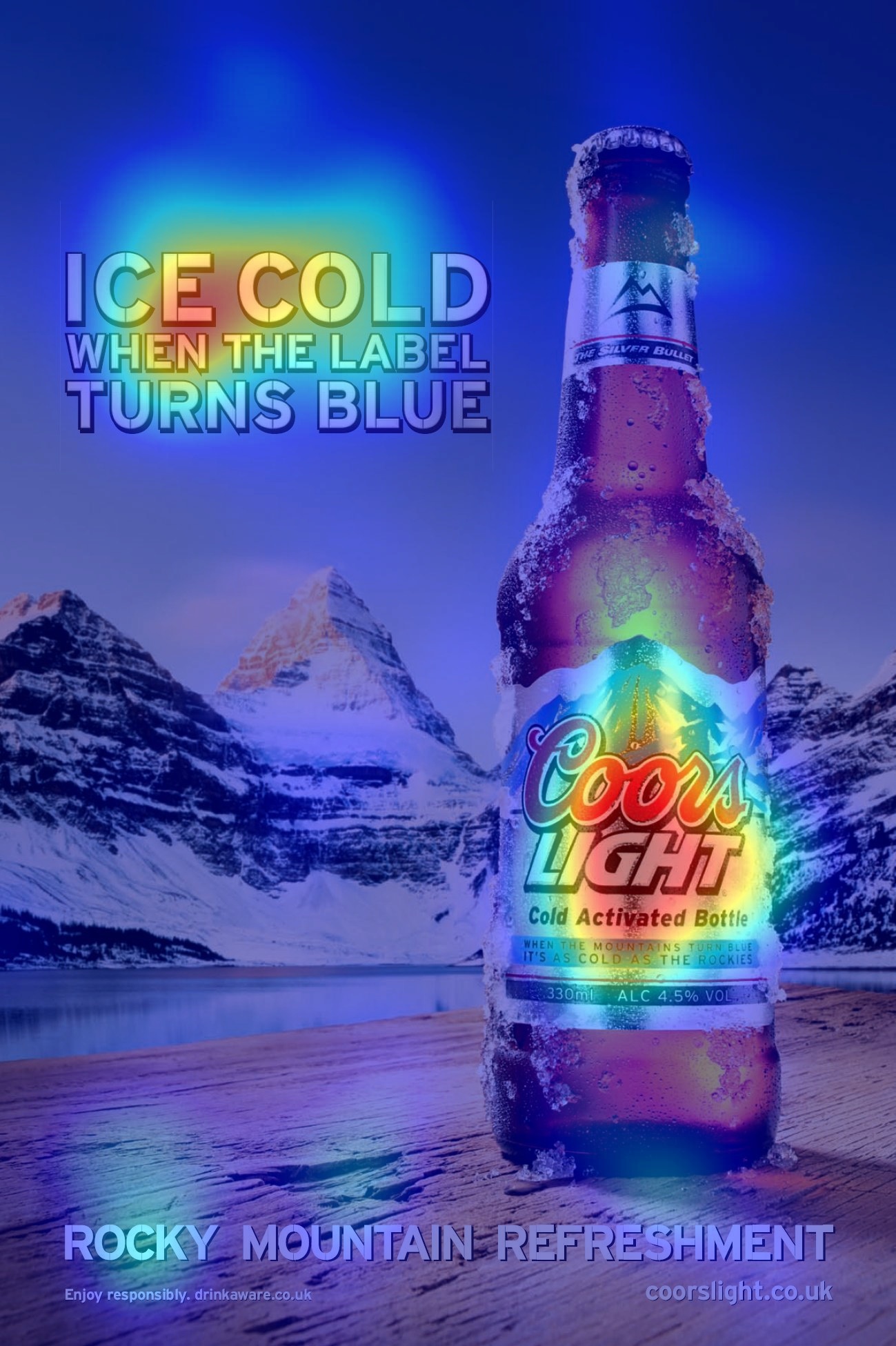} &
\includegraphics[width=0.11\linewidth]{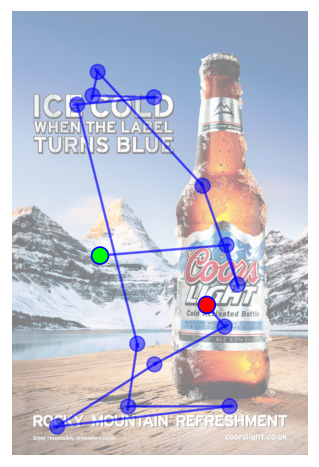} &
\includegraphics[width=0.11\linewidth]{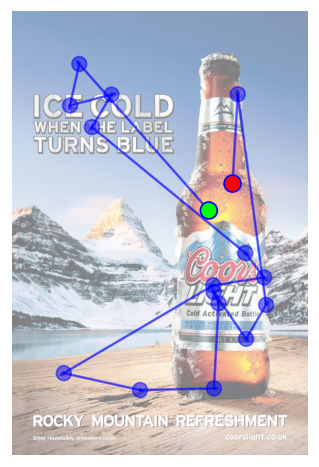} &
\includegraphics[width=0.11\linewidth]{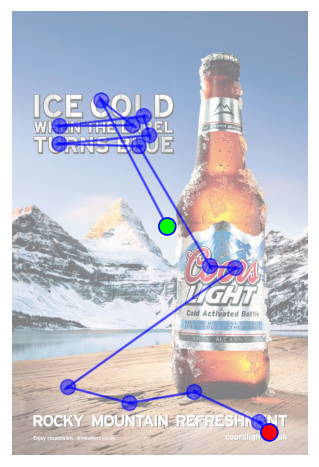} &
\includegraphics[width=0.11\linewidth]{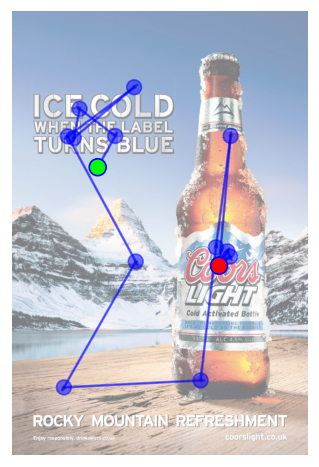}
\\
\end{tabular}
\caption{
    Examples of saliency maps and scanpaths in the UEyes dataset.
}
\label{fig:saliency_maps_scanpaths}
\end{figure*}

\paragraph{\textbf{Data Processing}}
We double-checked the collected data to guarantee the dataset's quality, and we removed any user data exhibiting inaccurate calibration or duplicate results.
Accordingly, the final dataset contains 94.86\% of the raw data collected. 
Fixations beyond image boundaries (6.8\% of the fixations) were not considered for analysis.

\subsection{Components of the Dataset}

In the following we describe the key features of our dataset.

\begin{description}

\item[Design images:] 495 images of each UI category, 1,980 images in total (55 blocks of 36 images each). 
.

\item[Eye-tracking logs:] 554 raw logs from the eye tracker in CSV format.
Each log includes eye movement data for one participant and one image block.

\item[Image types:] Categorization of each image in a separate CSV file.
Each image can belong to only one category.

\item[Multi-duration saliency maps:] Saliency maps for 1\,s, 3\,s, and 7\,s of free-viewing.
Each fixation is weighted by the time duration of each fixation.

\item[Scanpaths:] Sequences of fixations for participants. \autoref{fig:saliency_maps_scanpaths} shows examples of saliency maps and scanpaths.

\item[Segmentation information:] A JSON file for each UI with bounding boxes of detected images, texts, and faces. We modified the UIED model~\cite{xie2020uied} by (1)~solving a model limitation that ignored text detection
and (2)~integrating face detection with OpenCV face detection approach using Haar feature-based cascade classifiers~\cite{goyal2017face}.

\end{description}

%% file: 06-Discussion-and-Limitation.tex
  \setlength{\tabcolsep}{1pt}
  \def\arraystretch{1}%
 \begin{figure*}[h!]
 \def\w{0.175\linewidth}
 \def\ww{0.185\linewidth}
  \centering
 \begin{tabular}{l *4c}
 \bf \begin{turn}{90} 
 \bf  \ \ \ \   Image-oriented
 \end{turn} &
   \subfloat{{\includegraphics[height=\w]{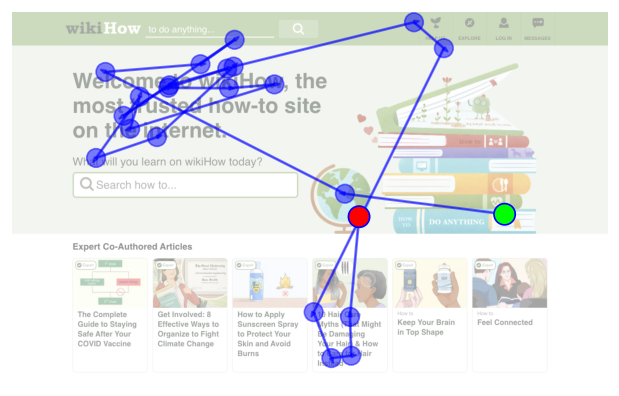}}} &
 \subfloat{{\includegraphics[height=\w]{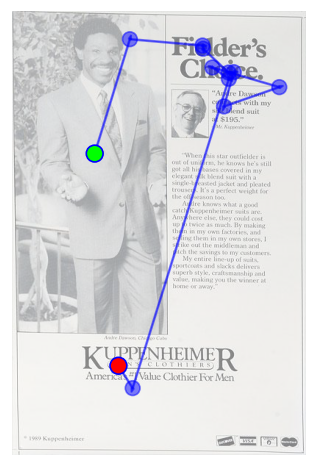}}} &
 \subfloat{{\includegraphics[height=\w]{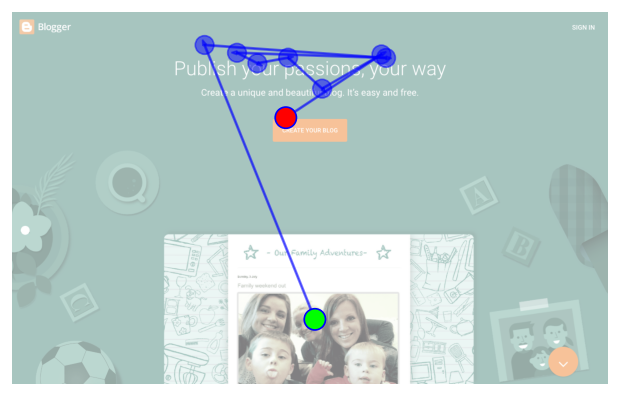}}} &
 \subfloat{{\includegraphics[height=\ww]{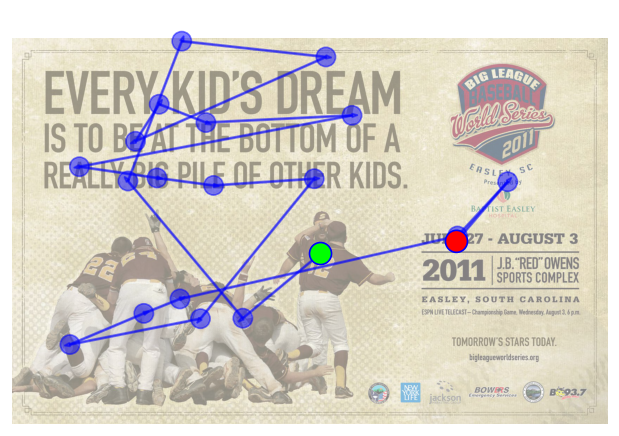}}} \\
 \bf \begin{turn}{90} 
 \bf \ \ \ \ \ \ Text-oriented
 \end{turn} &
   \subfloat{{\includegraphics[height=\w]{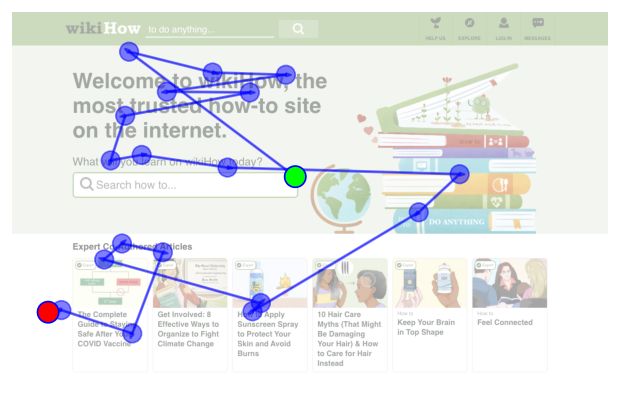}}} &
   \subfloat{{\includegraphics[height=\w]{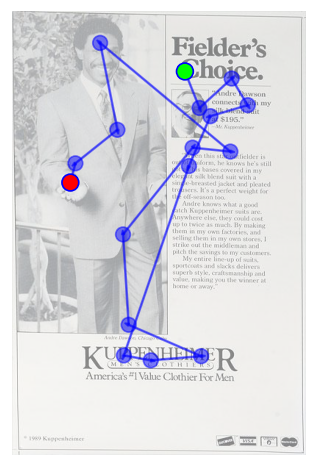}}} &
 \subfloat{{\includegraphics[height=\ww]{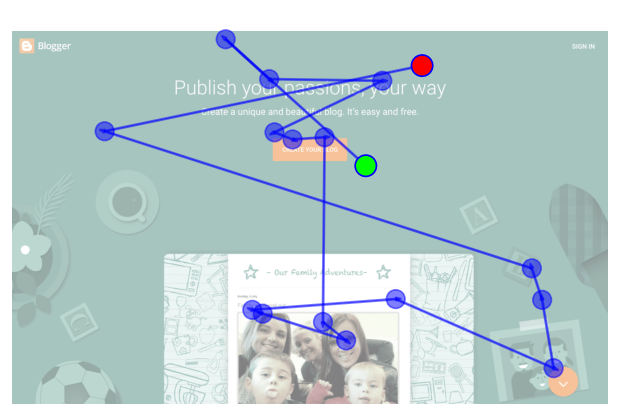}}}  &
   \subfloat{{\includegraphics[height=\w]{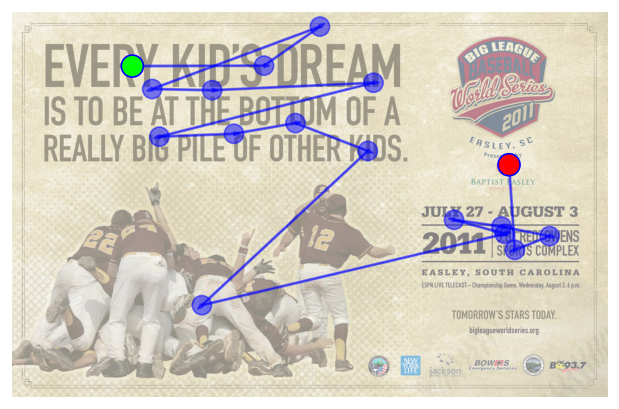}}} \\
 \end{tabular}
 \caption{
     Different users have different viewing strategies on user interfaces. 
     Image-oriented users often look at images before text, 
     while text-oriented users prefer the opposite.
 }
 \label{fig:user_scanpaths}
\end{figure*}

\section{How People Look at UIs}

We further analyzed our UEyes dataset and found that users tend to focus more on the upper-left region of a user interface, regardless of the type of UI, which is consistent with prior findings on mobile UIs~\cite{leiva2020understanding}. Additionally, we have observed that when users make saccades, their gaze tendsx to move towards the right or bottom portion of the UI. We have also found that movements towards the right tend to cover larger distances between consecutive points than movements towards the bottom.

In our analysis, we discovered that text elements are more likely to capture users' attention than images. This observation explains why saccades tend to move from left to right, rather than the opposite direction. It is worth noting, however, that this result could be due to our dataset, which primarily consisted of English-language interfaces that require left-to-right reading. Despite this, the ratio of images to text does not appear to influence the distribution of visited and revisited elements within these two categories. Additionally, we found that saccades towards the right-hand side of the interface tend to cover greater distances between consecutive points than those towards the bottom. This finding provides evidence that user interfaces are not viewed in the same way as natural scenes~\cite{leiva2020understanding}. Instead of a center bias, we observed a top-left bias. Our data allow a deeper dive into various subtle differences among the types of UIs examined. Several distinctions exemplify this:
\begin{description}

\item[Webpage:] 
When looking at webpages, participants showed a preference for scanning from left to right, resulting in larger distances between consecutive fixations compared to other types of user interfaces.
\item[Desktop UI:] Unlike in other types of user interfaces, fixations on desktop UIs are not evenly spread over the top-left quadrant. Instead, salient areas are divided into two regions: one right above the center and the other around the top-left corner.

\item[Mobile UI:] Compared to other types of user interfaces, mobile UIs have lower visit and revisit ratios. This suggests that users tend to concentrate more on a few elements of the UI that attract their attention while ignoring others. Moreover, there is less likelihood of revisiting the same elements.
\item[Poster:] Compared to desktop and mobile UIs, participants exhibited a much stronger tendency to scan from left to right, with only a small proportion of saccades directed from top to bottom. Additionally, the distances between consecutive fixation points showed more significant variation in this type of user interface than in others.

\end{description}

\section{Discussion and Future Work} 

Our study sheds new light on the eye-movement behavior that occurs with specific UI types. Here, we discuss potential future directions.

\subsection{Predictive Models}
The dataset can inform the assessment and improvement of computational models for visual saliency. Given a UI as input, a saliency model predicts saliency maps or scanpaths, which simulate how users perceive that UI. These models assist UI designers by predicting where users are likely to fix their gaze within a given design, enabling the designers to update it to emphasize the important areas better. Additionally, such models may help designers 'reflow' UI designs and create versions that maintain the desired visual emphasis across various screen sizes. In our recent work, we presented some improvements on predictive models \cite{jiang2023ueyes}. However, there is still ample room for improvement particularly in the accuracy of scanpath prediction.

\subsection{Individual Viewing Strategies}

We asked participants to self-report their viewing strategies. We found that some participants preferred to start by looking at images, while others focused on text or titles first. Participants also described different strategies for getting an overall idea of the UI. Some preferred to check the whole picture first, while others relied on titles to grab the general ideas and used images to understand the content better. Some participants indicated that they focused on the center of the page first, particularly when the layout was center-aligned, before scanning from top to bottom and left to right. \autoref{fig:user_scanpaths} show the different individual viewing strategies visually in our dataset. These findings provide insights into the various viewing strategies that individuals use when looking at UIs and can inform the development of more personalized predictive models.

To accurately predict gaze behavior, it is essential to consider the varying viewing strategies that individuals use when looking at user interfaces. Future research should focus on understanding and modeling these differences across UI types and at the individual level. Personalized predictive models that account for these differences can help improve the accuracy of gaze behavior predictions.

\subsection{Individual Element Influence}

further investigations can be conducted to explore the potential impact of images on the viewing path of a user interface (UI). For instance, if we replace the food image in the recipe app with a different image, such as a human face, would users still follow the same viewing path? It is worth noting that images have inherent visual saliency that is distinct from the UI itself, and replacing them may significantly alter user perception and viewing trajectory. To gain a deeper understanding of the effects of individual UI components, future research could aim to disentangle the effects of various elements, such as photos, text, and buttons, rather than considering the entire UI as an indivisible entity.

%% file: 07-Conclusion.tex
\subsection*{Open Science}
The dataset and trained models are available at \url{https://userinterfaces.aalto.fi/ueyeschi23}. %
\url{https://github.com/YueJiang-nj/UEyes-CHI2023}. 
The dataset includes raw CSV log files recorded with the GP3 HD eye tracker,
associated heatmaps and scanpaths, the image stimuli (screenshots), and metadata referring to the design type. This position paper has been presented at CHI2023 Computational UI Workshop~\cite{jiang2022computational, jiang2023future} and was presented in a demo at CHI2023 \cite{hegemann2023computational}.

%% file: main.bbl

\begin{thebibliography}{24}


\ifx \showCODEN    \undefined \def \showCODEN     #1{\unskip}     \fi
\ifx \showDOI      \undefined \def \showDOI       #1{#1}\fi
\ifx \showISBNx    \undefined \def \showISBNx     #1{\unskip}     \fi
\ifx \showISBNxiii \undefined \def \showISBNxiii  #1{\unskip}     \fi
\ifx \showISSN     \undefined \def \showISSN      #1{\unskip}     \fi
\ifx \showLCCN     \undefined \def \showLCCN      #1{\unskip}     \fi
\ifx \shownote     \undefined \def \shownote      #1{#1}          \fi
\ifx \showarticletitle \undefined \def \showarticletitle #1{#1}   \fi
\ifx \showURL      \undefined \def \showURL       {\relax}        \fi
\providecommand\bibfield[2]{#2}
\providecommand\bibinfo[2]{#2}
\providecommand\natexlab[1]{#1}
\providecommand\showeprint[2][]{arXiv:#2}

\bibitem[\protect\citeauthoryear{Bednarik and Tukiainen}{Bednarik and Tukiainen}{2007}]%
        {bednarik2007validating}
\bibfield{author}{\bibinfo{person}{Roman Bednarik} {and} \bibinfo{person}{Markku Tukiainen}.} \bibinfo{year}{2007}\natexlab{}.
\newblock \showarticletitle{Validating the restricted focus viewer: A study using eye-movement tracking}.
\newblock \bibinfo{journal}{\emph{Behavior research methods}} \bibinfo{volume}{39}, \bibinfo{number}{2} (\bibinfo{year}{2007}).
\newblock


\bibitem[\protect\citeauthoryear{Borji and Itti}{Borji and Itti}{2015}]%
        {borji2015cat2000}
\bibfield{author}{\bibinfo{person}{Ali Borji} {and} \bibinfo{person}{Laurent Itti}.} \bibinfo{year}{2015}\natexlab{}.
\newblock \showarticletitle{Cat2000: A large scale fixation dataset for boosting saliency research}.
\newblock \bibinfo{journal}{\emph{arXiv preprint arXiv:1505.03581}} (\bibinfo{year}{2015}).
\newblock


\bibitem[\protect\citeauthoryear{Borkin, Bylinskii, Kim, Bainbridge, Yeh, Borkin, Pfister, and Oliva}{Borkin et~al\mbox{.}}{2015}]%
        {borkin2015beyond}
\bibfield{author}{\bibinfo{person}{Michelle~A Borkin}, \bibinfo{person}{Zoya Bylinskii}, \bibinfo{person}{Nam~Wook Kim}, \bibinfo{person}{Constance~May Bainbridge}, \bibinfo{person}{Chelsea~S Yeh}, \bibinfo{person}{Daniel Borkin}, \bibinfo{person}{Hanspeter Pfister}, {and} \bibinfo{person}{Aude Oliva}.} \bibinfo{year}{2015}\natexlab{}.
\newblock \showarticletitle{Beyond memorability: Visualization recognition and recall}.
\newblock \bibinfo{journal}{\emph{IEEE transactions on visualization and computer graphics}} \bibinfo{volume}{22}, \bibinfo{number}{1} (\bibinfo{year}{2015}).
\newblock


\bibitem[\protect\citeauthoryear{Dataset}{Dataset}{2020}]%
        {desktopUI}
\bibfield{author}{\bibinfo{person}{Desktop~UI Dataset}.} \bibinfo{year}{2020}\natexlab{}.
\newblock \bibinfo{booktitle}{}.
\newblock
\urldef\tempurl%
\url{https://github.com/waltteri/desktop-ui-dataset}
\showURL{%
\tempurl}


\bibitem[\protect\citeauthoryear{Deka, Huang, Franzen, Hibschman, Afergan, Li, Nichols, and Kumar}{Deka et~al\mbox{.}}{2017}]%
        {deka2017rico}
\bibfield{author}{\bibinfo{person}{Biplab Deka}, \bibinfo{person}{Zifeng Huang}, \bibinfo{person}{Chad Franzen}, \bibinfo{person}{Joshua Hibschman}, \bibinfo{person}{Daniel Afergan}, \bibinfo{person}{Yang Li}, \bibinfo{person}{Jeffrey Nichols}, {and} \bibinfo{person}{Ranjitha Kumar}.} \bibinfo{year}{2017}\natexlab{}.
\newblock \showarticletitle{Rico: A Mobile App Dataset for Building Data-Driven Design Applications}. In \bibinfo{booktitle}{\emph{Proceedings of the 30th Annual Symposium on User Interface Software and Technology}} \emph{(\bibinfo{series}{UIST '17})}.
\newblock


\bibitem[\protect\citeauthoryear{Fosco, Casser, Bedi, O'Donovan, Hertzmann, and Bylinskii}{Fosco et~al\mbox{.}}{2020}]%
        {fosco2020predicting}
\bibfield{author}{\bibinfo{person}{Camilo Fosco}, \bibinfo{person}{Vincent Casser}, \bibinfo{person}{Amish~Kumar Bedi}, \bibinfo{person}{Peter O'Donovan}, \bibinfo{person}{Aaron Hertzmann}, {and} \bibinfo{person}{Zoya Bylinskii}.} \bibinfo{year}{2020}\natexlab{}.
\newblock \showarticletitle{Predicting visual importance across graphic design types}. In \bibinfo{booktitle}{\emph{Proceedings of the 33rd Annual ACM Symposium on User Interface Software and Technology}}. \bibinfo{pages}{249--260}.
\newblock


\bibitem[\protect\citeauthoryear{Goyal, Agarwal, and Kumar}{Goyal et~al\mbox{.}}{2017}]%
        {goyal2017face}
\bibfield{author}{\bibinfo{person}{Kruti Goyal}, \bibinfo{person}{Kartikey Agarwal}, {and} \bibinfo{person}{Rishi Kumar}.} \bibinfo{year}{2017}\natexlab{}.
\newblock \showarticletitle{Face detection and tracking: Using OpenCV}. In \bibinfo{booktitle}{\emph{2017 International conference of Electronics, Communication and Aerospace Technology (ICECA)}}, Vol.~\bibinfo{volume}{1}. IEEE, \bibinfo{pages}{474--478}.
\newblock


\bibitem[\protect\citeauthoryear{Hegemann, Jiang, Shin, Liao, Laine, and Oulasvirta}{Hegemann et~al\mbox{.}}{2023}]%
        {hegemann2023computational}
\bibfield{author}{\bibinfo{person}{Lena Hegemann}, \bibinfo{person}{Yue Jiang}, \bibinfo{person}{Joon~Gi Shin}, \bibinfo{person}{Yi-Chi Liao}, \bibinfo{person}{Markku Laine}, {and} \bibinfo{person}{Antti Oulasvirta}.} \bibinfo{year}{2023}\natexlab{}.
\newblock \showarticletitle{Computational Assistance for User Interface Design: Smarter Generation and Evaluation of Design Ideas}. In \bibinfo{booktitle}{\emph{Extended Abstracts of the 2023 CHI Conference on Human Factors in Computing Systems}}. \bibinfo{pages}{1--5}.
\newblock


\bibitem[\protect\citeauthoryear{Jiang, Huang, Duan, and Zhao}{Jiang et~al\mbox{.}}{2015}]%
        {jiang2015salicon}
\bibfield{author}{\bibinfo{person}{Ming Jiang}, \bibinfo{person}{Shengsheng Huang}, \bibinfo{person}{Juanyong Duan}, {and} \bibinfo{person}{Qi Zhao}.} \bibinfo{year}{2015}\natexlab{}.
\newblock \showarticletitle{SALICON: Saliency in Context}. In \bibinfo{booktitle}{\emph{2015 IEEE Conference on Computer Vision and Pattern Recognition (CVPR)}}. \bibinfo{pages}{1072--1080}.
\newblock
\urldef\tempurl%
\url{https://doi.org/10.1109/CVPR.2015.7298710}
\showDOI{\tempurl}


\bibitem[\protect\citeauthoryear{Jiang, Leiva, Houssel, Tavakoli, Kylmälä, and Oulasvirta}{Jiang et~al\mbox{.}}{2023a}]%
        {jiang2023ueyes}
\bibfield{author}{\bibinfo{person}{Yue Jiang}, \bibinfo{person}{Luis~A. Leiva}, \bibinfo{person}{Paul R.~B. Houssel}, \bibinfo{person}{Hamed~R. Tavakoli}, \bibinfo{person}{Julia Kylmälä}, {and} \bibinfo{person}{Antti Oulasvirta}.} \bibinfo{year}{2023}\natexlab{a}.
\newblock \showarticletitle{UEyes: Understanding Visual Saliency across User Interface Types}. In \bibinfo{booktitle}{\emph{Proceedings of the 2023 CHI Conference on Human Factors in Computing Systems}} (Hamburg, Germany) \emph{(\bibinfo{series}{CHI '23})}. \bibinfo{publisher}{Association for Computing Machinery}, \bibinfo{address}{New York, NY, USA}.
\newblock
\showISBNx{9781450394215}
\urldef\tempurl%
\url{https://doi.org/10.1145/3544548.3581096}
\showDOI{\tempurl}


\bibitem[\protect\citeauthoryear{Jiang, Lu, Lutteroth, Li, Nichols, and Stuerzlinger}{Jiang et~al\mbox{.}}{2023b}]%
        {jiang2023future}
\bibfield{author}{\bibinfo{person}{Yue Jiang}, \bibinfo{person}{Yuwen Lu}, \bibinfo{person}{Christof Lutteroth}, \bibinfo{person}{Toby Jia-Jun Li}, \bibinfo{person}{Jeffrey Nichols}, {and} \bibinfo{person}{Wolfgang Stuerzlinger}.} \bibinfo{year}{2023}\natexlab{b}.
\newblock \showarticletitle{The Future of Computational Approaches for Understanding and Adapting User Interfaces}. In \bibinfo{booktitle}{\emph{Extended Abstracts of the 2023 CHI Conference on Human Factors in Computing Systems}} (Hamburg, Germany) \emph{(\bibinfo{series}{CHI EA '23})}. \bibinfo{publisher}{Association for Computing Machinery}, \bibinfo{address}{New York, NY, USA}, Article \bibinfo{articleno}{367}, \bibinfo{numpages}{5}~pages.
\newblock
\showISBNx{9781450394222}
\urldef\tempurl%
\url{https://doi.org/10.1145/3544549.3573805}
\showDOI{\tempurl}


\bibitem[\protect\citeauthoryear{Jiang, Lu, Nichols, Stuerzlinger, Yu, Lutteroth, Li, Kumar, and Li}{Jiang et~al\mbox{.}}{2022}]%
        {jiang2022computational}
\bibfield{author}{\bibinfo{person}{Yue Jiang}, \bibinfo{person}{Yuwen Lu}, \bibinfo{person}{Jeffrey Nichols}, \bibinfo{person}{Wolfgang Stuerzlinger}, \bibinfo{person}{Chun Yu}, \bibinfo{person}{Christof Lutteroth}, \bibinfo{person}{Yang Li}, \bibinfo{person}{Ranjitha Kumar}, {and} \bibinfo{person}{Toby Jia-Jun Li}.} \bibinfo{year}{2022}\natexlab{}.
\newblock \showarticletitle{Computational Approaches for Understanding, Generating, and Adapting User Interfaces}. In \bibinfo{booktitle}{\emph{Extended Abstracts of the 2022 CHI Conference on Human Factors in Computing Systems}} (New Orleans, LA, USA) \emph{(\bibinfo{series}{CHI EA '22})}. \bibinfo{publisher}{Association for Computing Machinery}, \bibinfo{address}{New York, NY, USA}, Article \bibinfo{articleno}{74}, \bibinfo{numpages}{6}~pages.
\newblock
\showISBNx{9781450391566}
\urldef\tempurl%
\url{https://doi.org/10.1145/3491101.3504030}
\showDOI{\tempurl}


\bibitem[\protect\citeauthoryear{Judd, Durand, and Torralba}{Judd et~al\mbox{.}}{2012}]%
        {Judd2012benchmark}
\bibfield{author}{\bibinfo{person}{Tilke Judd}, \bibinfo{person}{Fr{\'e}do Durand}, {and} \bibinfo{person}{Antonio Torralba}.} \bibinfo{year}{2012}\natexlab{}.
\newblock \showarticletitle{A Benchmark of Computational Models of Saliency to Predict Human Fixations}. In \bibinfo{booktitle}{\emph{MIT Technical Report}}.
\newblock


\bibitem[\protect\citeauthoryear{Judd, Ehinger, Durand, and Torralba}{Judd et~al\mbox{.}}{2009}]%
        {judd2009learning}
\bibfield{author}{\bibinfo{person}{Tilke Judd}, \bibinfo{person}{Krista Ehinger}, \bibinfo{person}{Fr{\'e}do Durand}, {and} \bibinfo{person}{Antonio Torralba}.} \bibinfo{year}{2009}\natexlab{}.
\newblock \showarticletitle{Learning to predict where humans look}. In \bibinfo{booktitle}{\emph{2009 IEEE 12th international conference on computer vision}}. IEEE, \bibinfo{pages}{2106--2113}.
\newblock


\bibitem[\protect\citeauthoryear{Kim, Bylinskii, Borkin, Gajos, Oliva, Durand, and Pfister}{Kim et~al\mbox{.}}{2017}]%
        {kim2017bubbleview}
\bibfield{author}{\bibinfo{person}{Nam~Wook Kim}, \bibinfo{person}{Zoya Bylinskii}, \bibinfo{person}{Michelle~A Borkin}, \bibinfo{person}{Krzysztof~Z Gajos}, \bibinfo{person}{Aude Oliva}, \bibinfo{person}{Fr\~edo Durand}, {and} \bibinfo{person}{Hanspeter Pfister}.} \bibinfo{year}{2017}\natexlab{}.
\newblock \showarticletitle{BubbleView: an interface for crowdsourcing image importance maps and tracking visual attention}.
\newblock \bibinfo{journal}{\emph{ACM Transactions on Computer-Human Interaction (TOCHI)}} \bibinfo{volume}{24}, \bibinfo{number}{5} (\bibinfo{year}{2017}).
\newblock
\urldef\tempurl%
\url{https://doi.org/10.1145/3131275}
\showDOI{\tempurl}


\bibitem[\protect\citeauthoryear{Kim, Bylinskii, Borkin, Oliva, Gajos, and Pfister}{Kim et~al\mbox{.}}{2015}]%
        {kim2015crowdsourced}
\bibfield{author}{\bibinfo{person}{Nam~Wook Kim}, \bibinfo{person}{Zoya Bylinskii}, \bibinfo{person}{Michelle~A Borkin}, \bibinfo{person}{Aude Oliva}, \bibinfo{person}{Krzysztof~Z Gajos}, {and} \bibinfo{person}{Hanspeter Pfister}.} \bibinfo{year}{2015}\natexlab{}.
\newblock \showarticletitle{A crowdsourced alternative to eye-tracking for visualization understanding}. In \bibinfo{booktitle}{\emph{Proceedings of the 33rd Annual ACM Conference Extended Abstracts on Human Factors in Computing Systems}}. \bibinfo{pages}{1349--1354}.
\newblock


\bibitem[\protect\citeauthoryear{Leiva, Xue, Bansal, Tavakoli, K{\"o}ro{\dh}lu, Du, Dayama, and Oulasvirta}{Leiva et~al\mbox{.}}{2020}]%
        {leiva2020understanding}
\bibfield{author}{\bibinfo{person}{Luis~A Leiva}, \bibinfo{person}{Yunfei Xue}, \bibinfo{person}{Avya Bansal}, \bibinfo{person}{Hamed~R Tavakoli}, \bibinfo{person}{Tu{\dh}{\c{c}}e K{\"o}ro{\dh}lu}, \bibinfo{person}{Jingzhou Du}, \bibinfo{person}{Niraj~R Dayama}, {and} \bibinfo{person}{Antti Oulasvirta}.} \bibinfo{year}{2020}\natexlab{}.
\newblock \showarticletitle{Understanding visual saliency in mobile user interfaces}. In \bibinfo{booktitle}{\emph{22nd International conference on human-computer interaction with mobile devices and services}}. \bibinfo{pages}{1--12}.
\newblock


\bibitem[\protect\citeauthoryear{Miniukovich and Marchese}{Miniukovich and Marchese}{2020}]%
        {miniukovich2020relationship}
\bibfield{author}{\bibinfo{person}{Aliaksei Miniukovich} {and} \bibinfo{person}{Maurizio Marchese}.} \bibinfo{year}{2020}\natexlab{}.
\newblock \showarticletitle{Relationship between visual complexity and aesthetics of webpages}. In \bibinfo{booktitle}{\emph{Proceedings of the 2020 CHI Conference on Human Factors in Computing Systems}}. \bibinfo{pages}{1--13}.
\newblock


\bibitem[\protect\citeauthoryear{Ramanathan, Katti, Sebe, Kankanhalli, and Chua}{Ramanathan et~al\mbox{.}}{2010}]%
        {ramanathan2010eye}
\bibfield{author}{\bibinfo{person}{Subramanian Ramanathan}, \bibinfo{person}{Harish Katti}, \bibinfo{person}{Nicu Sebe}, \bibinfo{person}{Mohan Kankanhalli}, {and} \bibinfo{person}{Tat-Seng Chua}.} \bibinfo{year}{2010}\natexlab{}.
\newblock \showarticletitle{An eye fixation database for saliency detection in images}. In \bibinfo{booktitle}{\emph{European conference on computer vision}}. Springer, \bibinfo{pages}{30--43}.
\newblock


\bibitem[\protect\citeauthoryear{Rosenholtz, Dorai, and Freeman}{Rosenholtz et~al\mbox{.}}{2011}]%
        {Rosenholtz11}
\bibfield{author}{\bibinfo{person}{Ruth Rosenholtz}, \bibinfo{person}{Amal Dorai}, {and} \bibinfo{person}{Rosalind Freeman}.} \bibinfo{year}{2011}\natexlab{}.
\newblock \showarticletitle{Do Predictions of Visual Perception Aid Design?}
\newblock \bibinfo{journal}{\emph{ACM Trans. Appl. Percept.}} \bibinfo{volume}{8}, \bibinfo{number}{2} (\bibinfo{year}{2011}).
\newblock


\bibitem[\protect\citeauthoryear{Still and Masciocchi}{Still and Masciocchi}{2010}]%
        {Still10}
\bibfield{author}{\bibinfo{person}{Jeremiah~D. Still} {and} \bibinfo{person}{Christopher~M. Masciocchi}.} \bibinfo{year}{2010}\natexlab{}.
\newblock \showarticletitle{A Saliency Model Predicts Fixations in Web Interfaces}. In \bibinfo{booktitle}{\emph{Proc. MDDAUI Workshop}}.
\newblock


\bibitem[\protect\citeauthoryear{Websites}{Websites}{2022}]%
        {Alexa500}
\bibfield{author}{\bibinfo{person}{Alexa Top~500 Websites}.} \bibinfo{year}{2022}\natexlab{}.
\newblock \bibinfo{howpublished}{\url{https://www.expireddomains.net/alexa-top-websites/}}.
\newblock


\bibitem[\protect\citeauthoryear{Xie, Feng, Xing, Chen, and Chen}{Xie et~al\mbox{.}}{2020}]%
        {xie2020uied}
\bibfield{author}{\bibinfo{person}{Mulong Xie}, \bibinfo{person}{Sidong Feng}, \bibinfo{person}{Zhenchang Xing}, \bibinfo{person}{Jieshan Chen}, {and} \bibinfo{person}{Chunyang Chen}.} \bibinfo{year}{2020}\natexlab{}.
\newblock \showarticletitle{UIED: A Hybrid Tool for GUI Element Detection}. In \bibinfo{booktitle}{\emph{Proceedings of the 28th ACM Joint Meeting on European Software Engineering Conference and Symposium on the Foundations of Software Engineering}} (Virtual Event, USA) \emph{(\bibinfo{series}{ESEC/FSE 2020})}. \bibinfo{publisher}{Association for Computing Machinery}, \bibinfo{address}{New York, NY, USA}, \bibinfo{pages}{1655–1659}.
\newblock
\showISBNx{9781450370431}
\urldef\tempurl%
\url{https://doi.org/10.1145/3368089.3417940}
\showDOI{\tempurl}


\bibitem[\protect\citeauthoryear{Xu, Ehinger, Zhang, Finkelstein, Kulkarni, and Xiao}{Xu et~al\mbox{.}}{2015}]%
        {xu2015turkergaze}
\bibfield{author}{\bibinfo{person}{Pingmei Xu}, \bibinfo{person}{Krista~A Ehinger}, \bibinfo{person}{Yinda Zhang}, \bibinfo{person}{Adam Finkelstein}, \bibinfo{person}{Sanjeev~R Kulkarni}, {and} \bibinfo{person}{Jianxiong Xiao}.} \bibinfo{year}{2015}\natexlab{}.
\newblock \showarticletitle{Turkergaze: Crowdsourcing saliency with webcam based eye tracking}.
\newblock \bibinfo{journal}{\emph{arXiv preprint arXiv:1504.06755}} (\bibinfo{year}{2015}).
\newblock


\end{thebibliography}
